# Quantum-Based Combinatorial Optimization for Optimal Sensor Placement in Civil Structures

**Gabriel San Martín**[1], **MSc.**
Department of Civil & Environmental Engineering,
The B. John Garrick Institute for the Risk Sciences,
University of California, Los Angeles, CA 90095
gsanmartin@ucla.edu

**Enrique López Droguett, PhD.**
Department of Civil & Environmental Engineering,
The B. John Garrick Institute for the Risk Sciences,
University of California, Los Angeles, CA 90095
eald@ucla.edu

## Abstract

Over the last decade, concepts such as industry 4.0 and the Internet of Things (IoT) have contributed to the increase in the availability and affordability of sensing technology. In this context, Structural Health Monitoring (SHM) arises as an especially interesting field to integrate and develop these new sensing capabilities, given the criticality of structural application for human life and the elevated costs of manual monitoring. Due to the scale of structural systems, one of the main challenges when designing a modern monitoring system is the Optimal Sensor Placement (OSP) problem. The OSP problem is combinatorial in nature, making its exact solution infeasible in most practical cases, usually requiring the use of meta-heuristic optimization techniques. While approaches such as genetic algorithms (GA) have been able to produce significant results in many practical case studies, their ability to scale up to more complex structures is still an area of open research. This study proposes a novel quantum-based combinatorial optimization approach to solve the OSP problem approximately, within the context of SHM. For this purpose, a Quadratic Unconstrained Binary Optimization (QUBO) model formulation is developed, taking as a starting point the modal strain energy (MSE) of the structure. The framework is tested using numerical simulations of Warren truss bridges of varying scale. The results obtained using the proposed framework are compared against exhaustive search approaches to verify their performance. More importantly, a detailed discussion of the current limitations of the technology and the future paths of research in the area is presented to the reader.

## 1 Introduction

Modern Structural Health Monitoring (SHM) constitutes a powerful framework to detect, locate, and identify damage in a structural system [1]. At its core, it is a hybrid technique that makes use of both sensing data and physical models to correctly assess the health state of one or more structural components. While in older structures the sensing technology and data collection systems relied heavily on expert inputs via the execution of manual inspections, the current trend is towards developing intelligent SHM systems that can leverage the increasing affordability and availability of modern sensing technology to remotely monitor a structure. This shift in paradigm serves multiple purposes. First, it increases the safety of the structure's operation, decreasing the exposure of human beings to the dangerous maneuvers required to collect data. Second, it decreases the downtime usually required to perform manual inspections. Finally, it allows expert personnel to continuously monitor the structure, detecting relevant damage indicators early and, therefore, decreasing even further the downtime and costs incurred due to corrective maintenance activity. All these advantages result in a structural system that is more reliable, safe, and cost-efficient to operate.

Examples of the use of intelligent SHM methodologies are abundant in the literature. For example, Meruane et. al. [2] presents how diverse machine learning techniques can be used to identify impacts in aerospace structural components remotely. Feng et al. [3] presents a review on the use of computer vision for SHM of civil infrastructure, ranging from the identification of dynamic response to the detection of damage. Azimi et al. [4] presents a state of the art review on the usage of deep learning techniques for the damage identification and quantification in civil structures. Finally, on a more general note, Bao et al. [5] present a general review on the usage of data-driven approaches for SHM, focusing the discussion on future trends and opportunities.

[1] Corresponding author.

However, what all these techniques have in common is that they are all mostly data driven. Consequently, they make the underlying assumption that quality information can be retrieved from the structure using an appropriate sensor configuration. This is not a trivial assumption to make. The identification of changes in the dynamic properties of the structure, which are mostly conducive to identifying and locating anomalies and damage, strongly depends on where the information is captured, i.e., where the sensors are located. This situation gives rise to the problem commonly known as the Optimal Sensor Placement (OSP) problem. An OSP problem is usually formulated as a min-max optimization problem. On one hand, the goal is to capture enough information such that the posterior monitoring and diagnosis tasks can be fulfilled with the desired accuracy. On the other hand, due to sensor costs and operational restrictions, this level of captured information should be achieved using the minimum number of sensors possible. The latter restriction is important because when sensors are integrated into a structure, they become another component in the system, requiring their own maintenance schedule and reliability assessment.

In mathematical terms, the OSP problem can be classified as a discrete optimization problem, where the requirement is to find an optimal sensor layout within the structure in accordance with some metric of fitness. Obtaining the exact solution for this class of optimization problems is difficult because they usually present a combinatorial nature: the requirement is often to choose $m$ sensor positions out of $n$ candidate locations in the structure ($n >> m$). This characteristic makes the number of feasible sensor layouts to rapidly increase with the scale of the structural system. For example, a small structure comprised of 10 candidate locations and a budget of five sensors will have 252 feasible configurations that need to be evaluated, while a medium system composed of 100 candidate locations and a budget of 10 sensors will result in $1.73 \times 10^{13}$ possible combinations. This characteristic of the OSP problem renders the obtention of its exact solution an intractable problem in practical situations.

Currently, researchers tackle the OSP problem using a wide variety of solution strategies. Broadly, these algorithms can be divided into two classes: deterministic and stochastic algorithms. The first class corresponds to those algorithms that use deterministic rules to transverse the solution space in search of high-performance configurations. Stochastic algorithms, in change, use stochastic rules to transition between feasible solutions, allowing them to transverse a wider amount of the solution space, and often reach better solutions. Heuristic and meta-heuristic strategies are usually in this category. They have gained considerable popularity over the past decades in the field of OSP due to their relatively high efficiency in the treatment of larger problems. Genetic algorithms [6] and biology-based algorithms [7] are among the most relevant examples of these classes of solution strategy for the OSP problem.

Genetic algorithms work by initially setting a starting population of candidate solutions, which are iteratively combined and evaluated, only preserving for the next iteration those that fulfill a predetermined criterion of fitness. After a certain termination criterion is met, the best solution found is proposed as the near-optimal solution. A varied range of genetic algorithms has been used in the determination of OSP configurations. For example, Yi et. al. [8] proposed an improved genetic algorithm for the determination of sensor placement in a high-rise building in the north of China. More recently, Civera et. al. [9] presented a multi-objective genetic algorithm that takes into account the different damage scenarios that may affect structures, guiding the optimization toward those configurations that can easily detect and identify certain types of failures. Biology-based algorithms are designed to emulate certain aspects of animals, in general the foraging behavior of insects, in the aid of locating high performance solutions. A modification of the well-known ant-colony optimization algorithm [10], widely used in the solution of the traveling salesman problem, was used by Feng et. al. [11] to improve the accuracy of OSP for a transmission tower model, beating traditional genetic algorithms. Sun & Büyüköztürk [12] applied a different algorithm inspired by the social behavior of bees. Their approach is tested using one simulated structural model and two real-world examples. Finally, Gomes & Pereira [13] proposed a modification to the firefly algorithm to perform SHM in the fuselage of commercial airplanes. For a complete review of the algorithms used for solving the OSP problem, the reader is referred to [14].

However, when the scale of the structural system is large, and therefore the number of candidate sensor locations is high, researchers have had to rely on preprocessing techniques or simplifying assumptions over the structural numerical models to reduce the number of locations before solving the optimization problem [14]. For example, Sun et al. [12] proposed to solve the OSP problem using the bee colony algorithm by first reducing the size of the original structural model with the Iterated Improved Reduced System (IIRS) technique. With this approach, they are able to tackle the OSP problem in a case study concerning a structural building with 185 degrees of freedom (DOFs) by reducing it first to a smaller system with only 74 DOFs. Blachoswi et al. [15] proposed the use of convex relaxation and model reduction to transform a large OSP instance into a smaller, continuous one. With these two simplifications, they tackle a case study consisting of a transit bridge with 71,804 nodes, which after the model reduction step consists of a total number of candidate locations equal to 3,447. Moreover, Yang et al. [16] proposes the use of a multi-objective optimization approach to connect the selection of nodes in the model reduction stage with the election of optimal DOFs for sensor location. They validated this approach in a numerical case study involving a truss structure connecting two satellite bodies consisting of 117 candidate positions. What all these solution strategies have in common is the utilization of a simplification step over the original structural system, which likely places another penalization over the general accuracy of the solution approach.

Our interest in this paper is to explore a novel category of Quantum Computing algorithms that have the potential to not require these types of model-reduction steps. These techniques are currently divided into two main categories depending on the physical implementation of the quantum hardware in which they are executed. The first category uses Quantum Annealing (QA) [17] to solve quadratic unconstrained binary optimization (QUBO) problems. Based on the Quantum Adiabatic theorem, these algorithms first prepare a quantum system to encode a given optimization problem. Then, they evolve this system towards a lower energy state that represents a near optimum of the original optimization problem [17]. In contrast, the second category approximates the behavior of quantum annealing using gate-based quantum computers. In general, gate-based quantum computing is a more flexible hardware implementation than QA and, therefore, it has received more attention from mainstream media and technology companies such as IBM or Google. The main approach for solving combinatorial problems using gate-based quantum computers is through the Quantum Approximation Optimization Algorithm (QAOA) ([18], [19]), an approximation model of QA that can leverage the recent advances in gate-based quantum hardware. Both categories of quantum optimization solvers have already seen use in practical case studies. For example, Herman et. al. [20] present a survey on quantum computing for its application in finance, with a special emphasis on portfolio optimization techniques using both QA and QAOA. Similarly, Yarkoni et. al. [17] present a review of quantum annealing for industry applications, which includes solutions for scheduling and traffic flow optimization problems. In a closer context to SHM, Speziali et. al. [21] applied QA to find a near-optimal solution for the placement of pressure sensors in a water distribution network (WDN). The proposed approach first models the WDN as a mathematical graph, where the nodes are either tanks or junctions, and the edges are pipes connecting the different parts of the system. The OSP configuration is found by solving the minimum vertex covering problem, ensuring that the number of edges (pipelines) that are not monitored by an adjacent pressure sensor is minimized.

Following the trend of new applications that arise from this new field of research, this paper's main objective is to consider the capabilities and limitations of gate-based quantum optimization for the solution of the OSP problem in civil structures. For this, a novel Quadratic Unconstrained Binary Optimization (QUBO) model based on the Modal Strain Energy (MSE) criteria is proposed and approximately solved using the QOAO algorithm. For this purpose, we resort to a series of numerical experiments performed using quantum computing simulation software, that are ran directly on a desktop computer. Two Warren truss bridges of different scales are used as case studies for these experiments, in order to show how the proposed methodology scales with the size of the structure. This paper constitutes, to the best of the authors' knowledge, the first application of QAOA for the OSP problem in a structural engineering context.

This paper makes three primary contributions. First, it aims to present a clear introduction to quantum computing for the civil engineering community, focusing on the relevant subject of sensor placement optimization. Second, using the aforementioned case studies, it aims to present the state of quantum computing optimization for the field, stating its current limitations and outlining future research opportunities. Third, and final, it outlines future research paths that tackle the limitations we have found, to hopefully motivate the community to start exploring the different uses of quantum computing in their respective fields.

The paper is organized as follows. Section 2 presents a brief introduction to the OSP problem, along with the main metric used as the objective function in its solution. Section 3 introduces a self-contained gate-based quantum computing theoretical background, focusing the attention on those aspects that are relevant to the QAOA technique, which is presented in Section 4. Section 5 presents the QUBO model formulation for the OSP problem and relevant implementation details for its application on the QAOA model. Then, Section 6 introduces the two case studies used to compare the proposed approach with exhaustive search solution techniques. Section 7 concludes the paper by critically assessing the current capabilities of quantum-based optimization techniques and outlines future paths of research in the area. Finally, a preprint has previously been published in [22].

# 2 Background on Optimal Sensor Placement for Civil Structures

SHM strategies using vibration-based modeling (VBM) rely on the hypothesis that damage can be detected, located, and identified using as a proxy variable the changes in the dynamic properties of different structural components. The dynamic behavior in the linear regime of most structures can be mathematically represented as a system of $N_{DOF}$ differential equations, where $N_{DOF}$ is the number of degrees of freedom in the structure. Equation (1) represents this system in matrix notation.

$$\boldsymbol{M}\ddot{\vec{X}}(t) + \boldsymbol{C}\dot{\vec{X}}(t) + \boldsymbol{K}\vec{X}(t) = \vec{F}(t) \tag{1}$$

In equation (1), $\boldsymbol{M}$, $\boldsymbol{C}$, and $\boldsymbol{K}$ are the mass, damping, and stiffness matrices of the system, respectively, while $\vec{F}(t)$ is an arbitrary excitation function. The determination of $\boldsymbol{M}$, $\boldsymbol{C}$, and $\boldsymbol{K}$ is usually dependent on both experimental and physical modeling techniques. The natural frequencies and modal shape vectors of the system can be obtained by solving the eigenvalue problem portrayed in equation (2).

$$\boldsymbol{M}^{-1}\boldsymbol{K}\vec{X} = \omega_n^2\vec{X} \tag{2}$$

Once solved, the squared natural frequencies and modal shape vectors of the system can be identified as the eigenvalues and eigenvectors, respectively. The modal shape vectors store important information about the dynamic characteristics of the system, as they constitute a linearly independent base from which the movement of the structure can be reconstructed. For this reason, many of the strategies to optimally locate sensors in a structure are focused on capturing a large amount of this modal information. To this extent, a wide variety of objective functions for the OSP problem have been formulated in the literature. A common category of fitness criteria groups those algorithms based on the Fisher Information Matrix (FIM). For example, the Effective Independence (EI) approach, originally proposed by Kammer [23], uses both the FIM and the mode shape matrix to find the sensor locations that maximize the amount of linear independency in the measured modal vectors.

However, the exposition in this paper is centered around the Modal Strain Energy criteria (MSE) [24], as it allows for the interpretation of the OSP problem as a Quadratic Unconstrained Binary Optimization (QUBO) problem, making it suitable for its implementation in quantum-based optimization schemes. The reason for this will become evident in section 5, where the Quantum Approximate Optimization Algorithm (QAOA) is detailed.

**Modal Strain Energy (MSE)**

An energy-based objective function for the OSP problem can be formulated by maximizing the strain energy captured by the chosen sensor locations. This function is known as the Modal Strain Energy [24] and its mathematical form is depicted in equation (3).

$$MSE(S) = \sum_{i=1}^{N_{DOF}} \sum_{j=1}^{N_{DOF}} \sum_{p \in S} \sum_{q \in S} \left| \phi_{pi} k_{pq} \phi_{qj} \right| \tag{3}$$

where $\phi_{pi}$ is the component corresponding to the $p$-th DOF of the i-th modal shape vector and $k_{pq}$ is the $(p, q)$ component of the stiffness matrix. While the MSE metric does not assures that the sensor locations will produce partial modal shape vectors with a lower degree of linear dependence, it has been reported that the modal strain energy is correlated with higher signal-to-noise ratios in the sensor locations ([25]–[27]), favoring the identification of damage in structures affected by perturbations from the environment or the usage of the structure itself.

# 3 Background on Gate-Based Quantum Computing

This section presents an introduction to gate-based quantum computing, following a functional approach. First, we introduce what a quantum computer is by briefly describing the three operations it can perform. Then, we explain each one of the operations in great detail, prioritizing a mathematical and computational explanation instead of a physics-based one. We continue our exposition with a description of the quantum circuit model, a graphical representation of algorithms that a quantum computer can execute. Finally, we close this introduction with a brief discussion about quantum computing simulators, the tool used to run all the results presented in this paper.

A quantum computer is a physical machine designed to perform three main operations: (i) create and store quantum states, (ii) modify those quantum states, (iii) and measure those quantum states to extract classical information. By performing these three operations, a quantum computer is capable of ingesting, processing, and returning information to a user, effectively performing computation. Whether this novel computation paradigm presents advantages over traditional computational paradigms is currently a field of active research. However, there is strong theoretical evidence to suggest that given a quantum computer with enough capacity, quantum computing can produce algorithms that are vastly more efficient in certain tasks than traditional algorithms [28], [29].

In what follows, we describe in detail each of these operations. However, it is necessary to first present a brief explanation of the notation used in quantum computing, since it uses symbols traditionally used in contexts outside of those of SHM and OSP.

## 3.1 Bra-Ket Notation

Before presenting our introduction to this novel topic, it is necessary to introduce the *bra-ket* notation, heavily used in quantum computing. For the rest of this article, a complex column vector $\vec{\Psi}_1 \in \mathbb{C}^{N\times 1}$ will be represented using the ket notation as $|\Psi_1\rangle$. Similarly, a conjugate transpose vector will be denoted using the *bra* notation as $\vec{\Psi}_2^\dagger = \langle\Psi_2| \in \mathbb{C}^{1\times N}$ . The inner product between these two vectors is written according to equation (4):

$$\langle\Psi_2|\Psi_1\rangle = \vec{\Psi}_2 \cdot \vec{\Psi}_1 \in \mathbb{C} \tag{4}$$

Additionally, the tensor product between two complex vectors $|\Psi_1\rangle \in \mathbb{C}^{N_1\times 1}$ and $|\Psi_2\rangle \in \mathbb{C}^{N_2\times 1}$ can be written in short form using the notation shown in equation (5):

$$|\Psi_1\rangle|\Psi_2\rangle = |\Psi_1\Psi_2\rangle = \vec{\Psi}_1 \otimes \vec{\Psi}_2 \in \mathbb{C}^{N_1} \otimes \mathbb{C}^{N_2} \quad (5)$$

The bra-ket notation is extensively used in quantum computing to represent the quantum states and quantum gates.

### 3.2 Creation and Storage of Quantum States: Qubits and Systems of Multiple Qubits

The quantum computer, as a hardware component, is capable of creating and storing a quantum state. A quantum state, in simpler terms, is the state (at a certain moment in time) of a quantum system. Mathematically speaking, this quantum state is represented as a complex vector $|\Psi\rangle \in \mathbb{C}^{N\times 1}$ with unitary norm. The simplest possible quantum state corresponds to the special case $N = 2$. In this case, the quantum state is commonly known as a qubit, and it represents the minimum unit of information in quantum computing. Equation (6) depicts a qubit using the ket notation, as a two-dimensional complex vector:

$$|\psi\rangle = \begin{bmatrix} c_0 \\ c_1 \end{bmatrix} \quad (6)$$

where $\{c_0, c_1\} \in \mathbb{C}$. Every qubit $|\psi\rangle$ can be decomposed as $|\psi\rangle = c_0|0\rangle + c_1|1\rangle$, where $|0\rangle$ and $|1\rangle$ are the basis vectors, shown in equation (7):

$$|0\rangle = \begin{bmatrix} 1 \\ 0 \end{bmatrix};\ |1\rangle = \begin{bmatrix} 0 \\ 1 \end{bmatrix} \quad (7)$$

Because of this decomposition, an arbitrary qubit $|\psi\rangle$ is said to be in a *superposition* of states $|0\rangle$ and $|1\rangle$, controlled by the complex coefficients $c_0$ and $c_1$.

A qubit $|\psi\rangle$ can also be represented graphically using the *Bloch sphere*. In the Bloch sphere, every possible quantum state that a qubit may encode is mapped to a point in the surface of a sphere of unitary radius. For this, the complex coefficients $c_0$ and $c_1$ are first represented in polar form, following equation (8).

$$|\psi\rangle = c_0|0\rangle + c_1|1\rangle = r_0 e^{i\xi_0}|0\rangle + r_1 e^{i\xi_1}|1\rangle \quad (8)$$

where $\xi_0$ and $\xi_1$ are angles. Equation (8) shows that a qubit's state depends on four real coefficients, $r_0, r_1, \xi_0, \xi_1 \in \mathbb{R}$. However, it can be demonstrated [30] that using the restriction of the Bloch sphere unitary radius, we can reduce this set to only two coefficients, obtaining the qubit representation shown in equation (9):

$$|\psi\rangle = \cos\frac{\theta_1}{2}|0\rangle + e^{i\theta_2}\sin\frac{\theta_1}{2}\,|1\rangle \quad (9)$$

where $0 \le \theta_1 \le \pi$ and $0 \le \theta_2 \le 2\pi$ are interpreted as the polar and azimuthal angles in a unitary sphere. It is straightforward to notice that the poles of the sphere, when $\theta_1 = 0$ or $\theta_1 = \pi$, represent the basis states $|0\rangle$ or $|1\rangle$, respectively. Figure 1 shows the Bloch sphere representing an arbitrary qubit state.

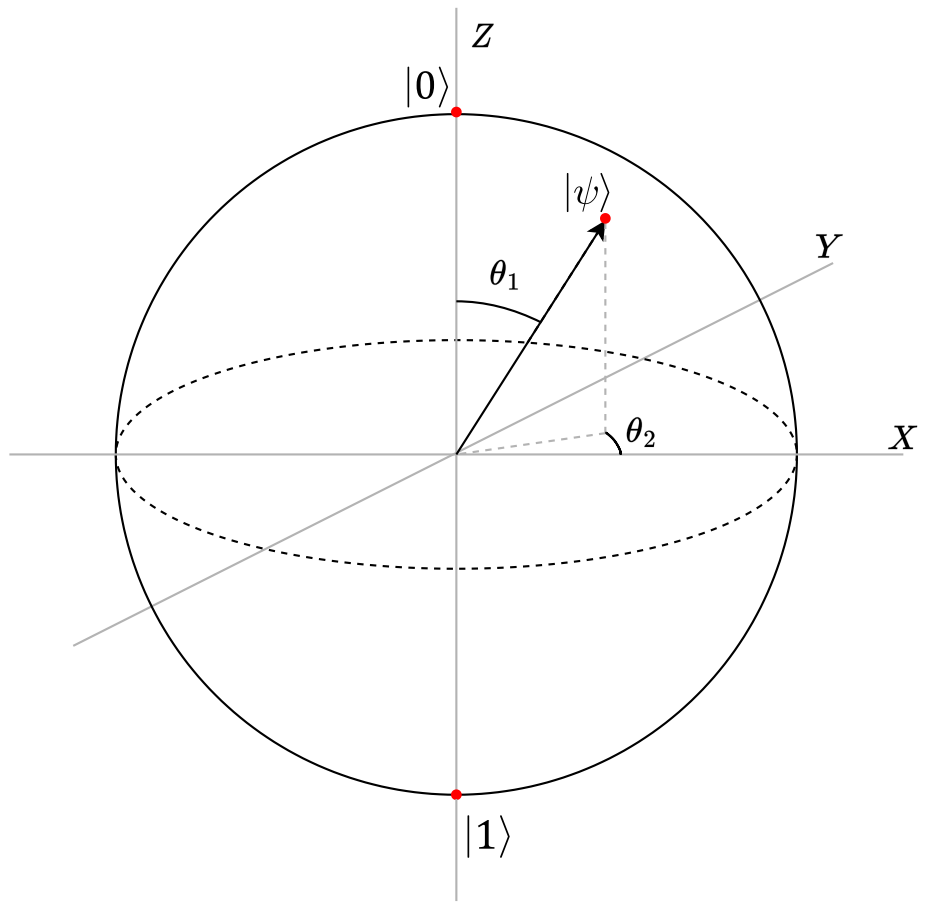

Figure 1: Bloch sphere representation of a single qubit. The surface of the unitary sphere encodes all the possible quantum states that a qubit may hold. As a result, quantum states can also be completely characterized, up to a global phase, by two real numbers: $\theta_1$ and $\theta_2$.

As it will be shown in section 3.3, the Bloch sphere allows for the interpretation of transformation operations over quantum states as rotations around the X, Y, and Z axis.

In a similar fashion to classical computing systems, where independent bits are of little use, single qubit systems are heavily limited in the computations they can carry. For this reason, quantum computers usually joined qubits together to generate useful quantum states. Given two qubits, $|\psi_1\rangle = c_{10}|0\rangle + c_{11}|1\rangle$ and $|\psi_2\rangle = c_{20}|0\rangle + c_{21}|1\rangle$, their multi-qubit system is described by the outer product presented in equation (10):

$$|\psi_1\psi_2\rangle = |\psi_1\rangle \otimes |\psi_2\rangle = \begin{bmatrix} c_{10}c_{20} \\ c_{10}c_{21} \\ c_{11}c_{20} \\ c_{11}c_{21} \end{bmatrix} \tag{10}$$

Equation (10) can also be expressed differently. Reinterpreting $c_1 = c_{10}c_{20}, c_2 = c_{10}c_{21}, c_3 = c_{11}c_{20}$, and $c_4 = c_{11}c_{21}$, the vector $|\psi_1\psi_2\rangle$ can be written as:

$$|\psi_1\psi_2\rangle = c_1\begin{bmatrix} 1 \\ 0 \\ 0 \\ 0 \end{bmatrix} + c_2\begin{bmatrix} 0 \\ 1 \\ 0 \\ 0 \end{bmatrix} + c_3\begin{bmatrix} 0 \\ 0 \\ 1 \\ 0 \end{bmatrix} + c_4\begin{bmatrix} 0 \\ 0 \\ 0 \\ 1 \end{bmatrix} = c_1|00\rangle + c_2|01\rangle + c_3|10\rangle + c_4|11\rangle \tag{11}$$

where the following short notation for the outer product has been used: $|0\rangle \otimes |0\rangle = |00\rangle$, $|0\rangle \otimes |1\rangle = |01\rangle$, $|1\rangle \otimes |0\rangle = |10\rangle$, and $|1\rangle \otimes |1\rangle = |11\rangle$. This pattern can be naturally extended to larger systems composed of more than two qubits. In general, an $n$-qubit system will consist of complex coefficients $\{c_i\}_{i=1}^{2^n}$, given by the elements of the vector $|\Psi\rangle = \otimes_{i=1}^{2^n}|\psi_i\rangle$. This manner of constructing quantum states in a quantum computer has the following two crucial implications. First, all practical quantum states in quantum computing will be of dimension $N = 2^n$, where $n$ is the number of qubits used (and combined) in the system. Second, and most importantly, there is an exponential scaling between the capacity of the quantum computer, measuring in its number of qubits, and the total number of complex coefficients that it can store. This particularity of quantum computing is the cornerstone of why researchers believe it may hold promise to tackle challenges that remain unresolved for traditional computers.

Another interesting quantum computing property is entanglement. Some quantum states are mathematically and physically valid, but they cannot be written as the Kronecker product of $n$ sub-states. An example of this is the state given in equation (12):

$$|\Psi\rangle = \frac{1}{\sqrt{2}}|00\rangle + \frac{1}{\sqrt{2}}|11\rangle \tag{12}$$

To see this, note that the system of equations presented in equation (13) does not have a solution. If neither $c_{10}$, $c_{20}$, $c_{11}$ or $c_{21}$ can be zero following the restriction that $c_1$ and $c_4$ are nonzero, then if follows that $c_2$ and $c_3$ should also be nonzero, which would not represent the quantum state depicted in equation (12):

$$\begin{aligned} c_1 &= c_{10}c_{20} = \frac{1}{\sqrt{2}} \\ c_2 &= c_{10}c_{21} = 0 \\ c_3 &= c_{11}c_{20} = 0 \\ c_4 &= c_{11}c_{21} = \frac{1}{\sqrt{2}} \end{aligned} \tag{13}$$

States that cannot be written as the Kronecker product of $n$ sub-states, such as the one presented in equation (12), are known as *entangled* states. On the other hand, quantum states that can be represented as a Kronecker product are known as separable states. Entanglement is a very important quantum computing property, as it allows for the creation of conditional relationships between qubits. For example, if the first qubit in the system depicted in equation (12) is found in the $|0\rangle$ state, then it automatically follows that the second qubit must also be in the $|0\rangle$ state (due to the complex coefficient accompanying $|01\rangle$ being zero). How to modify the quantum state of a system, and how to create entangled states in a quantum computer will be reviewed in section 3.3.

### 3.3 Modification of Quantum States: Quantum Gates and Unitary Matrices

Quantum states can be modified through the successive application of *quantum gates*. Quantum gates are linear maps represented by unitary[2] matrices $U$. Once a unitary operation is applied to a quantum state, a new quantum state is obtained. This state evolution is mathematically described in equation (14), where we have used $t$ as a variable indexing the steps in the overall transformation that a quantum state may go through in a quantum computer.

$$|\Psi_{t+1}\rangle = U|\Psi_t\rangle \tag{14}$$

Quantum gates are usually divided into single qubit gates and multi-qubit gates, according to the number of qubits they act on. Table 1 shows the most common single qubit gates, along with an interpretation of their effect on the Bloch's sphere.

Table 1: Single qubit gates and their matrix representation. An interpretation of the gates' effects on the Bloch's sphere is also included for the reader's convenience.

| Gate | Symbol | Matrix form | Bloch's sphere interpretation |
|---|---|---|---|
| Pauli-X | $P_X$ | $\begin{bmatrix} 0 & 1 \\ 1 & 0 \end{bmatrix}$ | Rotation by $\pi$ with respect to the X-axis. |

[2] A unitary matrix $U$ is a complex square matrix where its inverse is also its conjugate transpose: $U^\dagger U = UU^\dagger = I$

| | | | |
|---|---|---|---|
| Pauli-Y | $P_Y$ | $\begin{bmatrix} 0 & -i \\ i & 0 \end{bmatrix}$ | Rotation by $\pi$ with respect to the Y-axis. |
| Pauli-Z | $P_Z$ | $\begin{bmatrix} 1 & 0 \\ 0 & -1 \end{bmatrix}$ | Rotation by $\pi$ with respect to the Z-axis. |
| Rotation-X | $R_x$ | $\begin{bmatrix} \cos\left(\frac{\xi}{2}\right) & -i\sin\left(\frac{\xi}{2}\right) \\ -i\sin\left(\frac{\xi}{2}\right) & \cos\left(\frac{\xi}{2}\right) \end{bmatrix}$ | Rotation by $\xi$ with respect to the X-axis. |
| Rotation-Y | $R_y$ | $\begin{bmatrix} \cos\left(\frac{\xi}{2}\right) & -\sin\left(\frac{\xi}{2}\right) \\ \sin\left(\frac{\xi}{2}\right) & \cos\left(\frac{\xi}{2}\right) \end{bmatrix}$ | Rotation by $\xi$ with respect to the Y-axis. |
| Rotation-Z | $R_z$ | $\begin{bmatrix} e^{-i\frac{\xi}{2}} & 0 \\ 0 & e^{i\frac{\xi}{2}} \end{bmatrix}$ | Rotation by $\xi$ with respect to the Z-axis. |
| Hadamard | $Had$ | $\frac{1}{\sqrt{2}}\begin{bmatrix} 1 & 1 \\ 1 & -1 \end{bmatrix}$ | Rotation by $\pi$ with respect to the axis formed by the unitary vector with elevation angle $\frac{\pi}{4}$ and null azimuthal angle. |

As can be seen in Table 1, some quantum gates are parametric, i.e., they can receive an external parameter in the form of an angle $\xi$. These gates will be shown to be fundamental to encoding optimization problems into a quantum computer.

Multi-qubit gates are in general used to induce conditional relationships between the states of different qubits. The most used gates that represent these conditional relationships are the controlled-not (CX) gate and Toffoli (CCX) gate, also known as the double controlled-not gate. Both gates are portrayed in Table 2, along with their matrix form.

Table 2: multi-qubit gates and their matrix representation. Matrices dimensions and the number of qubits ($n$) are always related by the relationship $U \in \mathbb{C}^{2^n \times 2^n}$.

| **Gate** | **Symbol** | **Number of qubits, $n$** | **Matrix form** |
|---|---|---|---|
| Controlled -NOT | $CX$ | 2 | $\begin{bmatrix} 1 & 0 & 0 & 0 \\ 0 & 1 & 0 & 0 \\ 0 & 0 & 0 & 1 \\ 0 & 0 & 1 & 0 \end{bmatrix}$ |
| Toffoli Gate | $CCX$ | 3 | $\begin{bmatrix} 1 & 0 & 0 & 0 & 0 & 0 & 0 & 0 \\ 0 & 1 & 0 & 0 & 0 & 0 & 0 & 0 \\ 0 & 0 & 1 & 0 & 0 & 0 & 0 & 0 \\ 0 & 0 & 0 & 1 & 0 & 0 & 0 & 0 \\ 0 & 0 & 0 & 0 & 1 & 0 & 0 & 0 \\ 0 & 0 & 0 & 0 & 0 & 1 & 0 & 0 \\ 0 & 0 & 0 & 0 & 0 & 0 & 0 & 1 \\ 0 & 0 & 0 & 0 & 0 & 0 & 1 & 0 \end{bmatrix}$ |

The discussion around all the existing quantum gates and their effects over quantum states is large enough to escape the scope of this paper. For a complete review about quantum gates and what structures they can form, the reader is referred to [31].

In a similar fashion to qubits being concatenated to form larger quantum states, quantum gates can also be concatenated using the Kronecker product to generate a global unitary operation that conforms with the expression $|\Psi_{t+1}\rangle = U|\Psi_t\rangle$, where $|\Psi_{t+1}\rangle$and $|\Psi_t\rangle$ are the states before and after the application of the unitary matrix. This concatenation process is better explained through an example. Let us assume that a Hadamard gate, Rotation-Y gate, and CNOT gate are applied over a four-qubit system. This situation is depicted in Figure 2a. The final state before the measurement operation can be computed as $|\Psi\rangle = U|\psi_1\psi_2\psi_3\psi_4\rangle$, with $U$ given by the expression in equation (15).

$$U = Had \otimes R_y(\xi_1) \otimes CN \in \mathbb{C}^{16\times16} \tag{15}$$

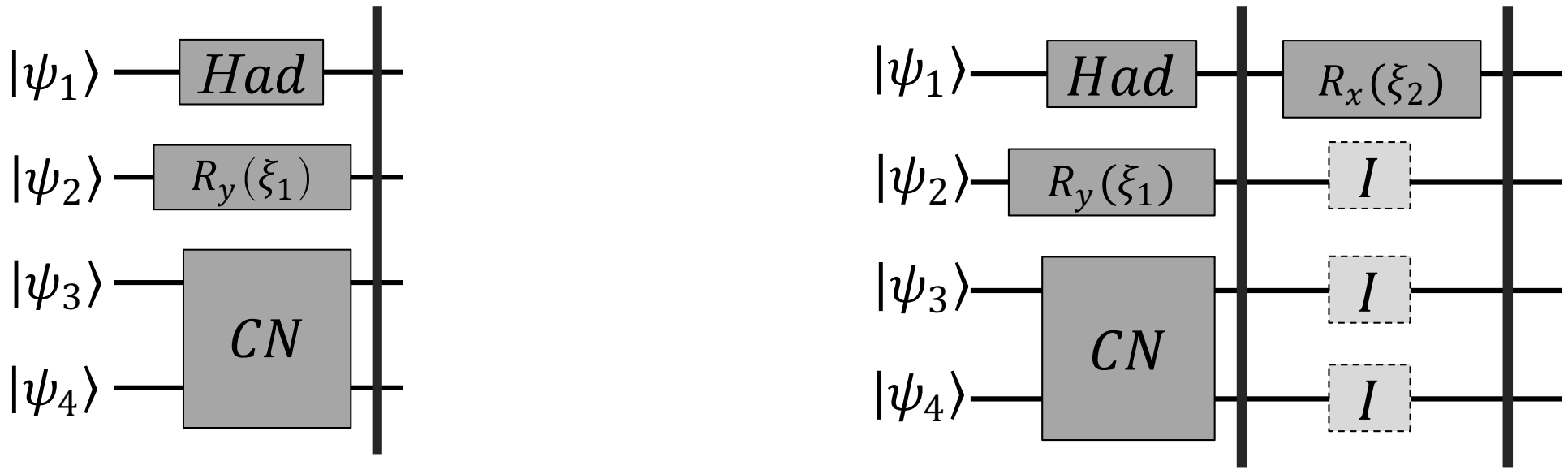

(a) The system does not require padding as every qubit is modified by the same number of gates.

(b) The system requires padding as qubit $|\psi_1\rangle$ is the only affected by a second round of quantum operations.

Figure 2: Quantum gates applied to multi-qubit systems. (a) shows a possible combination where single gates are applied to certain qubits, while a multi-qubit gate is applied to the sub-system composed by $|\psi_3\rangle$ and $|\psi_4\rangle$.(b) shows how padding with identity gates is used to complete the system of gates.

Figure 2b portrays a slightly different scenario, in which a Rotation-X gate has been applied over qubit $|\psi_1\rangle$ after the initial Hadamard gate. For this case, the final quantum state before the measurement operation can be computed as $|\Psi\rangle = U_2U_1|\psi_1\psi_2\psi_3\psi_4\rangle$, where $U_1$ is given by equation (15) and $U_2$ is formed by padding the missing gates in qubits $\psi_2$, $\psi_3$ and $\psi_4$ with identity gates, resulting in the expression shown in equation (16):

$$U_2 = R_x(\xi_2) \otimes I \otimes I \otimes I \in \mathbb{C}^{16\times16} \tag{16}$$

Padding with identity gates is necessary for $U_2$ to be an element in $\mathbb{C}^{16\times16}$ making the expression $|\Psi\rangle = U_2U_1|\psi_1\psi_2\psi_3\psi_4\rangle$,valid from a matrix multiplication perspective. Expanding this example to the general case of an $n$-qubit system, any operation over a quantum state can be understood as the application of $m$ independent unitary operations $\{U_i\}_i^m$ where each $U_i \in \mathbb{C}^{2^n\times2^n}$. Furthermore, the total operation is the result of multiplying the unitary operations, according to $U = \prod_{i=1}^m U_i$. This process of consolidation of unitary operations is shown graphically in Figure 3.

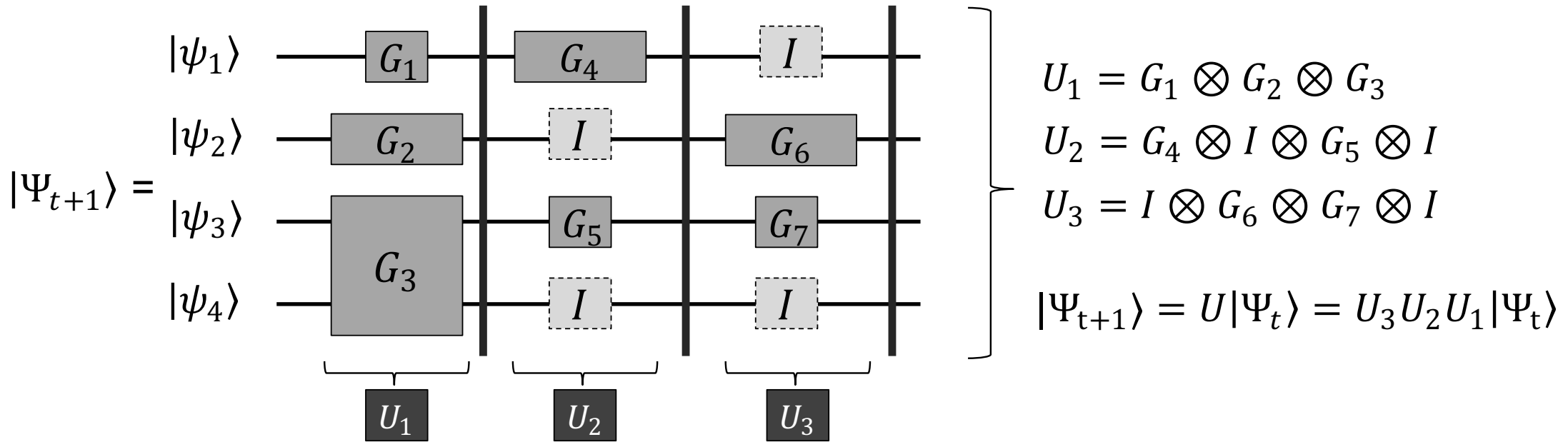

Figure 3: Consolidation of quantum operations. It is always possible to express the application of $m$ gates over an $n$-qubit system as one unitary operation by the usage of padding with unitary gates and the Kronecker product.

The dependance between the number of qubits that a system contains and the size of the matrices it can encode hints at the main advantage that quantum computing has over traditional computing. As an example, a quantum computer with 50 qubits can encode a unitary matrix $U \in \mathbb{C}^{2^{50} \times 2^{50}}$, which contains more than $1.267 \times 10^{30}$ complex-valued entries. Assuming that each entry is stored using 16 bytes, storing this matrix in a traditional computer would require roughly $2 \times 10^{16}$petabytes[3] of information. This amount of storage, and more importantly, the processing capabilities to operate such an exorbitantly large matrix is several orders of magnitude bigger than what is traditionally available to researchers and practitioners nowadays. As such, we can think of quantum computing as a technology that can enable algorithms based on complex-linear algebra that require the operation of extremely large constituents.

### 3.4 Measurement Operations Over the Quantum State

So far, we have reviewed how a quantum computer creates and stores a quantum state through the combination of basic units of information called qubits, and how these quantum states can be modified through the application of unitary matrices known as quantum gates. Now, we turn our attention to how information can be read out from the quantum computer.

Notably, quantum states cannot be observed directly. In other words, it is impossible to know exactly, and at the same time, the value of all the complex coefficients that compose the quantum state. This restriction is fundamental to quantum mechanics, and it cannot be overruled. However, what can be done is to perform a physical operation over the quantum state known as "measurement". In mathematical terms, the result of this operation performed over $|\Psi\rangle \in \mathbb{C}^{2^n}$ will be one its basis vectors $|e_i\rangle$, $i \in \{1, \dots, 2^n\}$, in accordance with the probability distribution described in equation (17).

$$p(|e_i\rangle) = \|c_i\|^2 \tag{17}$$

where $c_i$ is the $i$-th complex coefficient in the quantum state. In other words, the probability of obtaining a certain basis vector as a result of the measurement operation is completely controlled by the squared norm of the corresponding complex coefficient. Three important conclusions can be derived from equation (17). First, we can see that this presents a physics-based justification for the norm of all quantum states to be unitary: it is a necessary condition for the measurement probability distribution to be valid. Second, if we repeatedly construct and measure the quantum state, we can estimate the values of the squared norm of its entries; however, we will never be able to estimate the values of the original complex coefficients. Finally, and most importantly, a quantum state can be understood as an "object" encoding a categorical probability distribution over a space of $2^n$ elements. Moreover, the process of applying quantum gates to modify the quantum state can be interpreted as transforming this probability distribution.

[3] A petabyte is equal to 1,000,000 gigabytes.

As the quantum gates are unitary, and unitary matrices preserve the norm of the vector upon which they are applied, we are always assured that the resulting probability distribution will be a valid one.

Before continuing exploring what are the implications of this probabilistic interpretation of quantum computing, let us review some brief examples to clarify this measurement operation. Starting from the simplest possible case, equation (18) show the probability distribution corresponding to a single qubit $|\psi\rangle = [c_0,\ c_1]^T$.

$$P(|0\rangle) = \frac{\|c_0\|^2}{\|c_0\|^2 + \|c_1\|^2};\ P(|1\rangle) = \frac{\|c_1\|^2}{\|c_0\|^2 + \|c_1\|^2} \tag{18}$$

Extending the probability distribution presented in equation (18), measurement operations over two-qubits systems are presented in equation (19):

$$\begin{aligned} P(|00\rangle) &= \frac{|c_0|^2}{\sum_{i=0}^{3}|c_i|^2} \\ P(|01\rangle) &= \frac{|c_1|^2}{\sum_{i=0}^{3}|c_i|^2} \\ P(|10\rangle) &= \frac{|c_2|^2}{\sum_{i=0}^{3}|c_i|^2} \\ P(|11\rangle) &= \frac{|c_3|^2}{\sum_{i=0}^{3}|c_i|^2} \end{aligned} \tag{19}$$

More generally, a measurement operation is described as a Hermitian[4] matrix $H_v \in \mathbb{C}^{2^n \times 2^n}$. In this context, the sub-index $v = 1,2, \dots 2^n$ denotes one of the possible basis vectors of the system. That is, the measurement operation will measure the probability of obtaining one specific outcome. This probability, $p(|e_v\rangle)$ is given by equation (20):

$$p(|e_v\rangle) = \langle\Psi|H_v^\dagger H_m|\Psi\rangle \tag{20}$$

where $|\Psi\rangle$ is the quantum state just before the measurement operation is applied. While a variety of measurement operators exist, this overview will be focused on projective measurements. These types of operators are formed by performing the outer product between two vectors basis vectors.

For example, for a single qubit system $|\psi\rangle = c_0|0\rangle + c_1|1\rangle$, equations (21) and (22) depict the operators corresponding to the pure states $|0\rangle$ and $|1\rangle$:

$$H_{|0\rangle} = |0\rangle\langle 0| = \begin{pmatrix} 1 & 0 \\ 0 & 0 \end{pmatrix} \tag{21}$$

$$H_{|1\rangle} = |1\rangle\langle 1| = \begin{pmatrix} 0 & 0 \\ 0 & 1 \end{pmatrix} \tag{22}$$

If $H_{|0\rangle}$ is applied to $|\Psi\rangle$, then the probability of measuring $|0\rangle$ is given by equation (23), where the algebra involved has been omitted for the sake of brevity.

[4] A complex square Hermitian matrix $H$ fulfills $H = H^\dagger$ i.e., it is equal to its conjugate transpose.

$$p(|e_v\rangle = |0\rangle) = \langle\psi|H_{|0\rangle}^{\dagger}H_{|0\rangle}|\psi\rangle = \langle\psi||0\rangle\langle 0||0\rangle\langle 0||\psi\rangle = |c_0|^2 \quad (23)$$

Projective measurement operations extend naturally to multi-qubit systems, where the Hermitian matrices are also formed by applying an outer product operation between basis vectors. For example, in a three-qubit system, a possible measurement operation is $H_{|010\rangle} = |010\rangle\langle 010|$. This operation is used to measure the probability of obtaining as a final deterministic state the bitstring 010, or in physical terms, to find the first, second, and third qubits in the states 0, 1, and 0, respectively.

Another important measurement operation is to compute the expectation of a quantum state with respect to certain Hermitian operator $H_A$. This expectation is given by the expression presented in equation (24). As it will be shown in section 4.2, this operation is fundamental for computing the expected cost associated with the solution obtained from a combinatorial optimization algorithm.

$$\langle H_A\rangle_{|\Psi\rangle} = \langle\Psi|H_A|\Psi\rangle \quad (24)$$

The stochastic nature of the qubits (and in consequence, of quantum systems) is one of the key differences when compared to the deterministic behavior of classical *bits*. The use of complex coefficients to encode probability distributions allows the use of *interference* between quantum states. If probabilities were to be defined as real numbers, then such probabilities would only increase when added. Following the setting presented in quantum computing, as probability amplitudes are added, they can interfere with each other, effectively resulting in a lower overall probability for a particular outcome after taking the squared norm of the result. A physical interpretation of the inference property can be found in the way waves can interfere between them, amplifying or nullifying the resultant amplitude. Interference represents one of the pillars of gate-based quantum computing, where a set of pre-defined operations are applied to a quantum system to increase the likelihood of certain states that represent the solution for a computation problem.

### 3.5 Quantum Circuit Model

The framework presented in this section characterizes a quantum computer as a machine that creates, modifies, and measures quantum states. This process is done through the combination of different qubits to generate an exponentially larger quantum state, and the successive applications of unitary matrices to this quantum state. *Quantum programming*, or the act to "program" a quantum computer, can be summarized as selecting which gates to apply to the initial quantum state, and if any of those gates are parametric, which parameters do we enforce. This definition generates a quantum algorithm, which is no other thing than a simple set of instructions describing which gates, in which order, and with which parameters to apply over the initial quantum state. The initial quantum state of a quantum computer is traditionally initialized in the basal state $|00 \ldots 0\rangle$, or $|0\rangle^{\otimes n}$ for brevity. This gives all quantum algorithms the same starting point and standardizes their application.

The quantum algorithm can be graphically represented in a schematic known as *quantum circuit*. Figure 4 (center) depicts a diagram of a quantum circuit acting over $n$ qubits. The gates are represented by blocks that are applied over one or more qubits. At the end, we always include a measurement operation, describing that the quantum state is meant to be estimated through the process described in section 3.4.

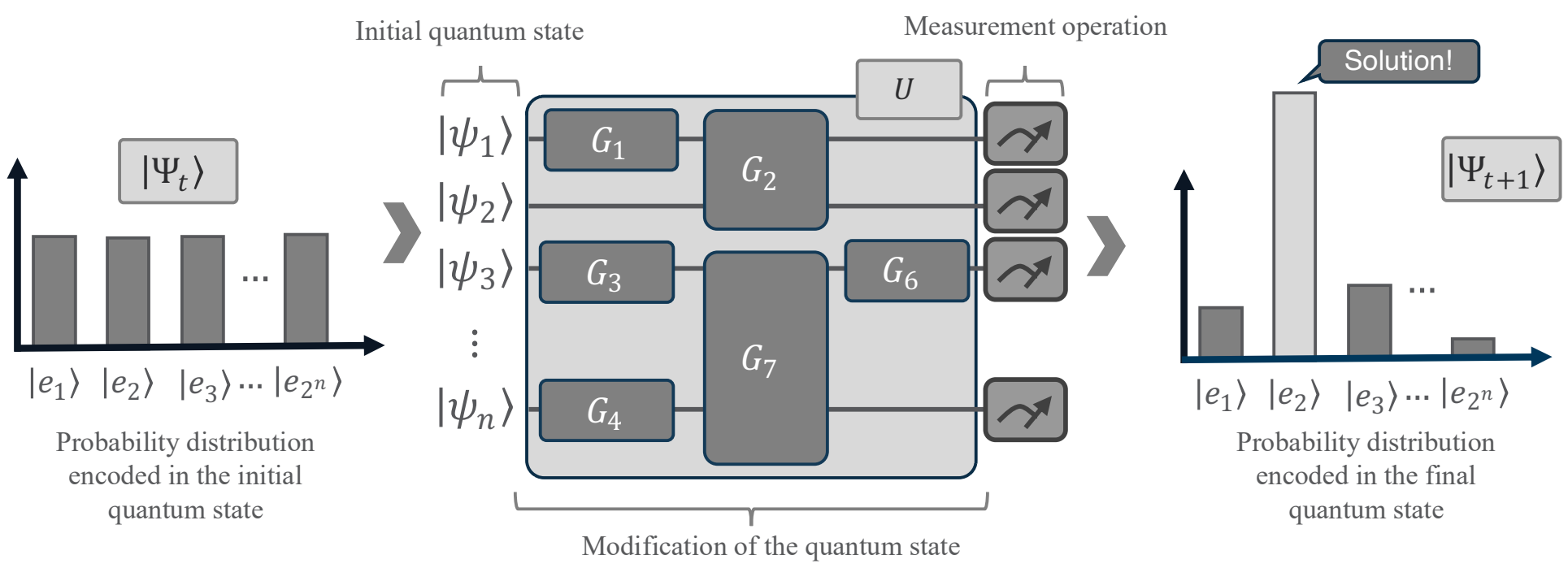

Figure 4: Quantum circuit model. To the left is the initial quantum state probability distribution. At the center, a series of single and multi-qubit gates are applied to the quantum states to transform it. Finally, to the right, the final quantum state's probability distribution is shown. The final objective of quantum computing is to enhance the probabilities of obtaining those basis states that represent the solution to a certain task encoded in the quantum circuit.

In terms of hardware, the current landscape of quantum computing is known as the Noisy Intermediate-Scale Quantum era (NISQ) [32]. NISQ hardware is characterized by having small to medium computing capacities (measured in the number of qubits available) and very limited error-correction capabilities. The practical effect of the latter is that the result of the estimation of the square norm of the entries of the quantum states are highly contaminated by noise in current quantum computers. In this context, the most used framework for researching quantum computing is using a quantum simulator. A quantum simulator is a specialized piece of software that runs on a traditional computer and simulates exactly, and without measurement errors, the process that occurs inside a quantum computer. That way, researchers can test quantum algorithms and obtain the same results that they would obtain on an error-corrected quantum computer, which is still in its early development stages. For this paper, all the results we will show are computed using a quantum simulator, the detailed specifications of which are fully stated in section 6.

While using a quantum simulator has clear advantages over current quantum hardware, and in many cases, it is the only approach to test quantum algorithms, it is not without its limitations. First, the traditional computer running the quantum simulator needs to store and operate the unitary matrices, which scale exponentially with the number of qubits. Because of this, simulating anything over 20 qubits requires High Performance Computing capabilities, even with efficient algorithms for matrix multiplication. Second, since the results of a quantum simulation are executing in a traditional computer, performance comparison are made difficult since time is no longer a valid option for metrics.

# 4 Quantum Approximate Optimization Algorithm for Combinatorial Optimization

A Combinatorial Optimization Problem (COP) is formulated as finding the optimal element among a discrete collection in accordance with a certain criterion of performance. A special class of COPs are binary optimization problems, where the candidate solutions are binary vectors, and the performance criterion is a mathematical function that takes the candidate solutions as inputs and returns a numerical score as the output. The OSP problem can be formulated as a binary optimization problem by considering $\{\vec{x}_i\}_{i=1}^{2^{N_{DOF}}}$ to be the set of candidate sensor configurations, represented as binary-valued vectors. If for given candidate configuration $\vec{x}_i$, the position $j$ is equal to 1 ($x_{ij} = 1$), then a sensor should be placed in degree of freedom $j$.

In general, the solution space of binary COPs is prohibitively large due to the number of possible combinations that can be formed. For this reason, an exhaustive search approach is usually infeasible for large-scale structures. In addition, given the discrete nature of the solution space, binary COPs cannot, in principle, be solved using gradient-based optimization techniques, such as Stochastic Gradient Descent. Because of this, solution methodologies for binary COPs often are designed as heuristics or meta-heuristics: algorithms that do not guarantee the discovery of an optimal solution, but that attempt to find a near-optimal solution in a reasonable computational time.

In this context, one of the most promising applications of quantum computing is as a meta-heuristic approach to solve binary COPs. The motivation for this is based on two aspects. First, as seen in section 3.5, repeatedly measuring a quantum system naturally results in a probability distribution over set of binary vectors. Additionally, this probability distribution can be modified by applying quantum gates. Consequently, finding the solution of a binary COP can be understood, in quantum computing terms, as finding a quantum circuit which when executed, assigns a high probability to the vector representing the optimal solution. Second, a quantum computer is at its core a physical system and, therefore, it can naturally evolve toward a lower energy state in the search for equilibrium. If a 1-to-1 mapping between the energy levels of the quantum system and the objective function value of each feasible solution can be established, then a relationship between solving the optimization problem and obtaining a lower energy state for the quantum system can be formulated. In what follows, the theoretical principles behind quantum-enhanced optimization are presented.

### 4.1 Quantum Time Evolution

The core concept behind quantum-enhanced optimization methods is the quantum adiabatic theorem. This theorem states that if a quantum system A is first prepared in one of its ground energy states and then transformed into quantum system B, then the final state will correspond to a ground energy state B if and only if the transition A→B was performed slow enough [33], [34]. Mathematically, this transition is given by equation (25), where $s(t)$ is a smooth transition function that fulfills $s(t=0)=0$ and $s(t=T)=1$, and $H_A$ and $H_B$ are Hermitian matrices known as Hamiltonians, representing the energy levels of quantum systems:

$$H(t) = \big(1 - s(t)\big)H_A + s(t)H_B \tag{25}$$

In equation (25), $H(t)$ represent the Hamiltonian of the combined quantum system at time $t$. The way in which a quantum state changes $|\Psi(t)\rangle$ through time depends on its energy (i.e., its Hamiltonian and is given by the Schrödinger equation, presented in equation (26):

$$i\frac{d}{dt}|\Psi(t)\rangle = H(t)|\Psi(t)\rangle \tag{26}$$

The solution of Schrödinger equation is shown in equation (27). Note how the transformation of a quantum state from 0 to $T$ is controlled by a matrix exponential operator.

$$|\Psi(T)\rangle = e^{-\int_0^T iH(t)dt}|\Psi(0)\rangle = U(0,T)|\Psi(0)\rangle \tag{27}$$

This matrix exponential operator is a unitary operator, and as such is commonly referred to as $U(0,T)$. The composition property of time-evolution operators in quantum mechanics dictates that if $t_1 < t_2$, then performing an evolution $U(0,t_2)$ is equivalent as performing two disjoint and successive evolutions $U(0,t_1)$ and $U(t_1,t_2)$. Using this property, equation (27) can be rewritten as:

$$\begin{aligned}|\Psi(T)\rangle &= U(T-\Delta t,T)U(T-2\Delta t,T-\Delta t)\dots U(\Delta t,2\Delta t)U(0,\Delta t)|\Psi(0)\rangle \\ &= \left(\prod_{j=1}^{P} U\big((j-1)\Delta t, j\Delta t\big)\right)|\Psi(0)\rangle\end{aligned} \tag{28}$$

where the application order of the unitary operators does matter (i.e., they are not commutative) and $\Delta t = T/P$. Each of the unitary operators $U_j$ are given by the expression presented in equation (29):

$$U_j = U\big((j-1)\Delta t, j\Delta t\big) = \exp\left(\int_{(\mathrm{j}-1)\Delta\mathrm{t}}^{j\Delta t} -iH(t)dt\right) \tag{29}$$

If $\Delta t$ is sufficiently small (or equivalently, $P$ is sufficiently high), then $H(t)$ can be considered as a constant in each interval and therefore the integral can be approximated according to equation (30):

$$U_j \approx \exp\big(-i\Delta t H(j\Delta t)\big) \tag{30}$$

Replacing $H(t)$ by the expression given in equation (25) provides the final form for the operator $U_j$:

$$U_j \approx \exp\Big(-i\Delta t\ \Big(\big(1 - s(j\Delta t)\big)H_A + s(j\Delta t)H_B\Big)\Big) \tag{31}$$

Equation (31) can be further divided using the properties of the exponential function only when the matrices $H_A$ and $H_B$ commute. Regrettably, that is not generally the case and therefore it is necessary to use the Lie-Trotter-Suzuki decomposition formula [35], which states that $\exp(i(H_1 + H_2)t)$ can be approximated by $\exp(iH_1 t)\exp(iH_2 t) + O(t^2)$. Ignoring higher order terms, equation (31) can be approximated by:

$$U_j \approx \exp\big(-i\Delta t(1 - s(j\Delta t)H_A)\big)\exp(-i\Delta t\ s(j\Delta t)H_B) \tag{32}$$

Finally, the evolution of the quantum state $|\Psi(t)\rangle$ from $t = 0$ to $t = T$ can be expressed by equation (33):

$$|\Psi(T)\rangle \approx \left(\prod_{j=1}^{P} \exp\big(-i\Delta t\big(1 - s(j\Delta t)\big)H_A\big)\exp(-i\Delta t\ s(j\Delta t)H_B)\right)|\Psi(0)\rangle \tag{33}$$

Equation (33) condenses the approach used by quantum computing to solve COPs. Let us suppose that $|\Psi(0)\rangle$ is defined as a ground energy level of the Hamiltonian $H_A$, and that $H_B$ is a Hamiltonian encoding the optimization problem of interest. Then, if the transformation between $H_A$ and $H_B$ is performed in multiple steps, i.e., $P \gg 1$, or equivalently $\Delta t \ll 1,$ then state $|\Psi(T)\rangle$ should represent a state that when measured will sample more frequently solutions vectors with a higher performance in terms of their objective value. Furthermore, the matrix exponentiation of a Hermitian matrix is a unitary matrix, and as such, can be recognized as a valid quantum gate. This allows us to rewrite equation (33) as:

$$|\Psi(T)\rangle \approx \left(\prod_{j=1}^{P} U_{H_A}(\beta_j)U_{H_B}(\gamma_j)\right)|\Psi(0)\rangle \tag{34}$$

where $\beta_j = \Delta t\big(1 - s(j\Delta t)\big)$ and $\gamma_j = \Delta t s(j\Delta t)$ are sets of parameters for the unitary operations.

However, it is still necessary to overcome two relevant challenges to fully realize an approach to solve a combinatorial optimization problem in a quantum computer: (i) how can we inform the quantum algorithm of the optimization problem? In other words, how can we encode the objective function of the optimization problem into a Hamiltonian matrix $H_B$? (ii) how can we represent the evolution of Hamiltonian $H_A$ and $H_B$ as a series of basic quantum gates in a gate-based quantum computer architecture. The following sections explain how to tackle these two challenges.

## 4.2 Hamiltonian Encoding of QUBO Problems

The objective is to encode a combinatorial optimization problem into a quantum system by mapping it to its total energy level (Hamiltonian). In the remainder of this subsection, the discussion will be limited to quadratic unconstraint binary optimization (QUBO) problems. Section 4.4 will detail how to incorporate constraints within the context of quantum computing optimization. The canonical form of QUBO problems applied to the OSP context is shown in equation (35):

$$\min_{\vec{x}\,\in\{\vec{x}_i\}_{i=1}^{2^{N_{DOF}}}} f(\vec{x}) = \vec{x}^T Q_1\,\vec{x} + Q_2\,\vec{x} \tag{35}$$

where $Q_1 \in \mathbb{R}^{2^{N_{DOF}}} \times \mathbb{R}^{2^{N_{DOF}}}$, $Q_2 \in \mathbb{R}^{2^{N_{DOF}}} \times \mathbb{R}^{2^{N_{DOF}}}$, and $\vec{x}$ is a vector representing a bitstring of size $2^{N_{DOF}}$.

According to Farhi et. al [19], the Hamiltonian operator $H$ encoding the QUBO problem should fulfill the following condition: given a measured quantum state of the form $|x\rangle = [0\ 0\ 0\ 0\ \dots\ 1\ \dots\ 0\ 0]^T \in \mathbb{R}^{2^{N_{DOF}}}$, where the 1 is located at position $i$, the result of computing the expectation of a Hamiltonian matrix $H$ under this state should be equal to the original cost function evaluated at the corresponding candidate sensor configuration $\vec{x}$. It is important to note that $|x\rangle$ is a representation of $\vec{x}$, but they are not equal. While $|x\rangle$ is a ket vector of $2^{N_{DOF}}$ dimensions that only contains one entry equal to one, $\vec{x}$ is a feasible solution vector of the original QUBO, and therefore is a vector of dimension $N_{DOF}$ that may contain any number of 1s and 0s among its entries. The relationship between $\vec{x}$ and $|x\rangle$ is bijective and given by:

$$|x\rangle = \bigotimes_{i=1}^{N_{DOF}} (1 - x_i)|0\rangle + x_i|1\rangle \tag{36}$$

where $x_i$ is the i-th element of the vector $\vec{x}$ and the $\otimes$ operator represents the Kronecker product. Take as an example the decision variable vector $\vec{x} = [1\ \ 1]^T$. Its corresponding ket vector is $|1\rangle \otimes |1\rangle = [0\ 0\ 0\ 1]^T$. According to equation (24), the expectation of Hamiltonian matrix $H$ given the state $|x\rangle$ is given by the expression presented in equation (37):

$$\langle H\rangle_{|x\rangle} = \langle x|H|x\rangle = H_{i,i} = f(\vec{x}) \tag{37}$$

From equation (37) is clear that $H$ should be a diagonal operator containing in its diagonal the corresponding value of the cost function for each possible bitstring of length $N_{DOF}$. As this matrix directly encodes the cost function of the optimization problem, it corresponds to the previous definition of $H_B$. Figure 5 presents a graphical depiction of $H_B$.

$$\vec{x}_1$$

$$H_B = \begin{bmatrix} f([0\ 0\ \dots 0\ 0]) & 0 & \dots & 0 \\ 0 & f([1\ 0\ \dots 0\ 0]) & \dots & 0 \\ \dots & \dots & \dots & \dots \\ 0 & 0 & \dots & f([1\ 1\ \dots 1\ 1]) \end{bmatrix} \in \mathbb{R}^{2^{N_{DOF}}} \times \mathbb{R}^{2^{N_{DOF}}}$$

$$\vec{x}_{2^{N_{DOF}}}$$

Figure 5: Cost Hamiltonian diagonal structure. As the optimization problem is identified as a QUBO problem, for $N_{DOF}$ decision variables the total number of possible candidate configurations is $2^{N_{DOF}}$.

However, directly constructing $H_B$ for a given relevant QUBO problem is in general infeasible. First, its size increases exponentially with the number of decision variables in the original problem. Second, an explicit construction involves computing the cost function for every possible solution vector, an approach that is equivalent to solving the QUBO problem by exhaustive search. Consequently, $H_B$ needs to be modeled implicitly. For this, Hadfield [36] proposed a series of rules to translate terms from a QUBO problem into the corresponding Hamiltonian form using varied combinations of Pauli-Z gates. The rules used in this paper are summarized in Table 3.

Table 3: Basic Boolean functions and their Hamiltonian representation assuming a QUBO problem with $N_{DOF}$ decision variables. The matrix $\bar{I}$ symbolizes an identity operator of size $2^{N_{DOF}} \times 2^{N_{DOF}}$. $P_{Z_i}$ is the result of $I \otimes I \otimes ... P_Z ... \otimes I$, a Kronecker product of $N_{DOF}$ terms where the i-th term is a Pauli-Z gate. Similarly, $P_{ZZ_{ij}}$ is the result of $I \otimes I \otimes ... P_Z ... \otimes ... P_Z \otimes I$, a Kronecker product of $N_{DOF}$ terms where the i-th and j-th terms are Pauli-Z gates.

| Terms in QUBO formulation | Hamiltonian Representation | QUBO Interpretation |
|---|---|---|
| Single binary term, $x_i$ | $\frac{1}{2}\bar{I} - \frac{1}{2}P_{Z_i}$ | Linear Term |
| Multiplication of two binary terms, $x_i \cdot x_j$ | $\frac{1}{4}\bar{I} - \frac{1}{4}\left(P_{Z_i} + P_{Z_j} - P_{ZZ_{ij}}\right)$ | Quadratic Term |

For example, the Hamiltonian corresponding to the two-variable cost function $f(x_1, x_2) = a_1x_1 + a_2x_2$ is given by equation (38).

$$f(x_1, x_2) = a_1x_1 + a_2x_2 \rightarrow H_B = a_1\left(\frac{1}{2}\bar{I} - \frac{1}{2}P_{Z_1}\right) + a_2\left(\frac{1}{2}\bar{I} - \frac{1}{2}P_{Z_2}\right) = \begin{bmatrix} 0 & 0 & 0 & 0 \\ 0 & a_2 & 0 & 0 \\ 0 & 0 & a_1 & 0 \\ 0 & 0 & 0 & a_1 + a_2 \end{bmatrix} \tag{38}$$

where the solution vector $\vec{x} = [0\ 0]^T$, corresponding to the ket vector $|x\rangle = |00\rangle = [1\ 0\ 0\ 0\ ]^T$ would produce an expectation $\langle 00|H_B|00\rangle = 0$, in accordance with the corresponding value of the cost function. As a second example, if $\vec{x} = [0\ 1]^T$, then $|x\rangle = |01\rangle = [0\ 1\ 0\ 0\ ]^T$ and the expectation results in $a_2$, again matching the value of the objective function at that point. The same behavior can be expected for the cases $\vec{x} = [1\ 0]^T$ and $\vec{x} = [1\ 1]^T$.

If the function is modified with a quadratic term, $f(x_1, x_2) = a_{12}x_1x_2 + a_1x_1 + a_2x_2$, the new Hamiltonian (according to Table 3) will be:

$$\begin{aligned} f(x_1, x_2) &= a_{12}x_1x_2 + a_1x_1 + a_2x_2 \\ \rightarrow H_2 &= a_{12}\left(\frac{1}{4}\bar{I} - \frac{1}{4}\left(P_{Z_1} + P_{Z_2} - P_{ZZ_{12}}\right)\right) + a_1\left(\frac{1}{2}\bar{I} - \frac{1}{2}P_{Z_1}\right) + a_2\left(\frac{1}{2}\bar{I} - \frac{1}{2}P_{Z_2}\right) \\ &= \begin{bmatrix} 0 & 0 & 0 & 0 \\ 0 & a_2 & 0 & 0 \\ 0 & 0 & a_1 & 0 \\ 0 & 0 & 0 & a_1 + a_2 + a_{12} \end{bmatrix} \end{aligned} \tag{39}$$

From equation (39), it is easy to note that the only state where the term $a_{12}$ is included into the cost function is the state $|x\rangle = [0\ 0\ 0\ 1]^T$, which correspond to $\vec{x} = [x_1\ x_2]^T = [1\ 1]$, as expected. A special case for the quadratic term is $x_ix_i$, which is equivalent to $x_i$ given that all variables are binary.

From the examples shown above, it is clear that the encoding of a cost function into a Hamiltonian following the rules presented in Table 3 is a linear operation. With this, a grouping strategy can be followed to implicitly represent any QUBO problem as a sum of three distinct types of Hamiltonians multiplied by real coefficients, according to equation (40):

$$H_B = a\bar{I} + \sum_{j=1}^{N} b_j P_{Z_j} + \sum_{j=1}^{N}\sum_{k=1}^{N} c_{jk}\, P_{ZZ_{jk}} \tag{40}$$

where $a$, $b_j$, and $c_{jk}$ are real constants. This approach is valid for any QUBO problem, independent of the size or number of cross-relationships between the variables.

### 4.3 Quantum Optimization Approximation Algorithm

The QAOA approach, originally proposed by Farhi et. al. [19], provides the practical implementation of equation (34) for a gate-based quantum computer. First, let us recall from section 4.1 that the evolution from the Hamiltonian $H_A$ towards the Hamiltonian $H_B$ can be approximated by the successive application of the parametric unitary matrices $U_{H_1}$ and $U_{H_2}$. Following the adiabatic theorem, the idea is to choose $H_A$ such that a ground energy state (eigenvector corresponding to the lower eigenvalue of the Hamiltonian matrix) is easily defined and prepared in a quantum circuit. For this, Farhi et. al. proposes the use of the following Hamiltonian:

$$H_A = \sum_{j=1}^{N_{DOF}} P_{X\,j} \tag{41}$$

where $P_{X\,j} = I \otimes I \otimes \dots P_X \dots \otimes I$ and $P_X$ is a Pauli-X operator located in the $j$-th position while the rest of the operators are identity gates. As expected, the operator $H_A \in \mathbb{R}^{2^{N_{DOF}}} \times \mathbb{R}^{2^{N_{DOF}}}$. The reasons for using this Hamiltonian are three-fold. First, a ground energy state can be easily generated by initializing an $N_{DOF}$-qubit system in the $|0\rangle^{\otimes N_{DOF}}$ state and then apply a Hadamard gate to each qubit of the system. Second, a Hamiltonian of this form will automatically not commute with the cost Hamiltonian $H_B$, a condition necessary for the QAOA algorithm to prevent it from becoming trapped in low-quality minima. Third, this Hamiltonian can be easily transformed into a unitary operation $U$ by performing matrix exponentiation. Given that the Pauli-X operator commutes with itself, it is possible to separate the exponentiation without using the Lie-Trotter-Suzuki decomposition formula, and write:

$$U_{H_A}(\beta_j) = \exp(-i\beta_j H_A) = e^{-i\beta_j \sum_{k=1}^{N_{DOF}} P_{X_k}} = \prod_{k=1}^{N_{DOF}} e^{-i\beta_j P_{X_k}} \tag{42}$$

Replacing this expression in equation (42), it is possible to obtain:

$$U_{H_A}(\beta_j) = \prod_{k=1}^{N_{DOF}} e^{-i\beta_j P_{X_k}} = \prod_{k=1}^{N_{DOF}} I \otimes I \otimes \dots \otimes e^{-i\beta_j P_X} \otimes \dots \otimes I \tag{43}$$

It can be proved that the expression $e^{-i\beta_j P_X}$ represents a unitary matrix corresponding to a Rotation-X quantum gate, according to equation (44).

$$e^{-i\beta_j P_X} = R_x(2\beta_j) \tag{44}$$

Equation (43) and Equation (44) show that the unitary operation corresponding to the Hamiltonian $H_A$ can be represented in a quantum circuit by applying to each qubit a Rotation-X gate controlled by an angle $2\beta_j = 2\Delta t(1 - s(j\Delta t))$. This circuit implementation is depicted in Figure 6.

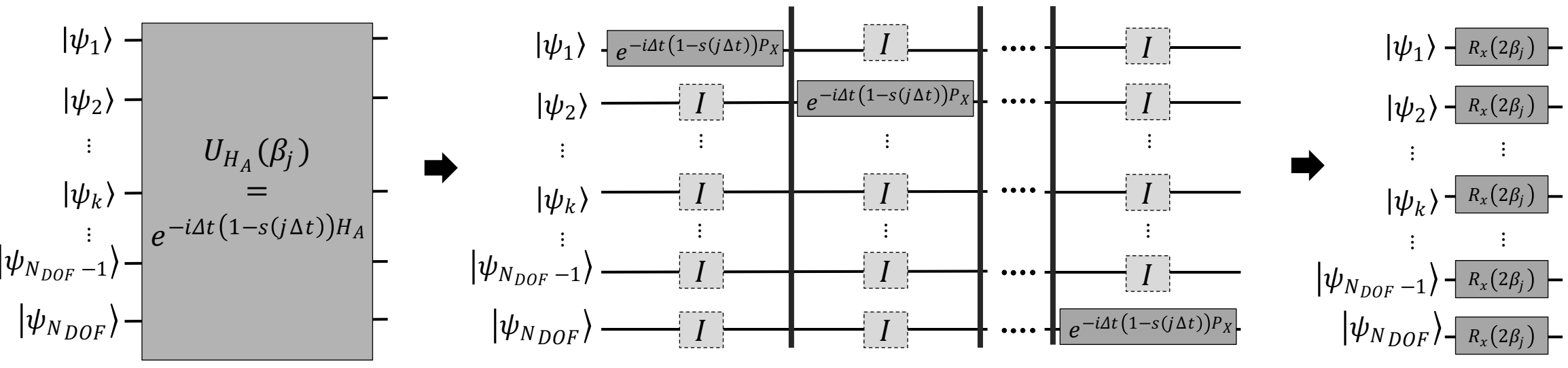

Figure 6: Circuit mapping for Hamiltonian $H_{\mathrm{A}}$.

The process of mapping the Hamiltonian $H_B$ into a unitary operation $U_{H_B}$ is similar. Combining equations (33) and (40), it follows that:

$$U_{H_B}(\gamma_j) = \exp\left(-i\gamma_j H_B\right) = \exp\left(-i\gamma_j\left(a\bar{I} + \sum_{k=1}^{N_{DOF}} b_k P_{Z_k} + \sum_{k=1}^{N_{DOF}}\sum_{m=1}^{N_{DOF}} c_{km}\, P_{ZZ_{km}}\right)\right) \tag{45}$$

It can be demonstrated that every term in $H_B$ commutes with each other and, therefore, equation (45) can be rewritten as:

$$U_{H_B}(\gamma_j) = \exp\left(-i\gamma_j a\bar{I}\right) \prod_{k=1}^{N_{DOF}} \exp\left(-i\gamma_j b_k P_{Z_k}\right) \prod_{k=1}^{N_{DOF}}\prod_{m=1}^{N_{DOF}} \exp\left(-i\gamma_j c_{km} P_{ZZ_{km}}\right) \tag{46}$$

The first term in equation (46) can be disregarded given that it is a constant overall phase and therefore it is canceled during the normalization process of the quantum state. As a result, the unitary operation $U_{H_B}$ consists of two types of terms: single and double Pauli-Z operators. Following a similar approach to that used in the mapping of $H_A$, the Pauli-Z Hamiltonian results in a Rotation-Z when mapped into a quantum circuit. In a similar fashion, the double Pauli-Z operator results in a combination of CNOT and Rotation-Z gates. These mappings are shown in Figure 7.

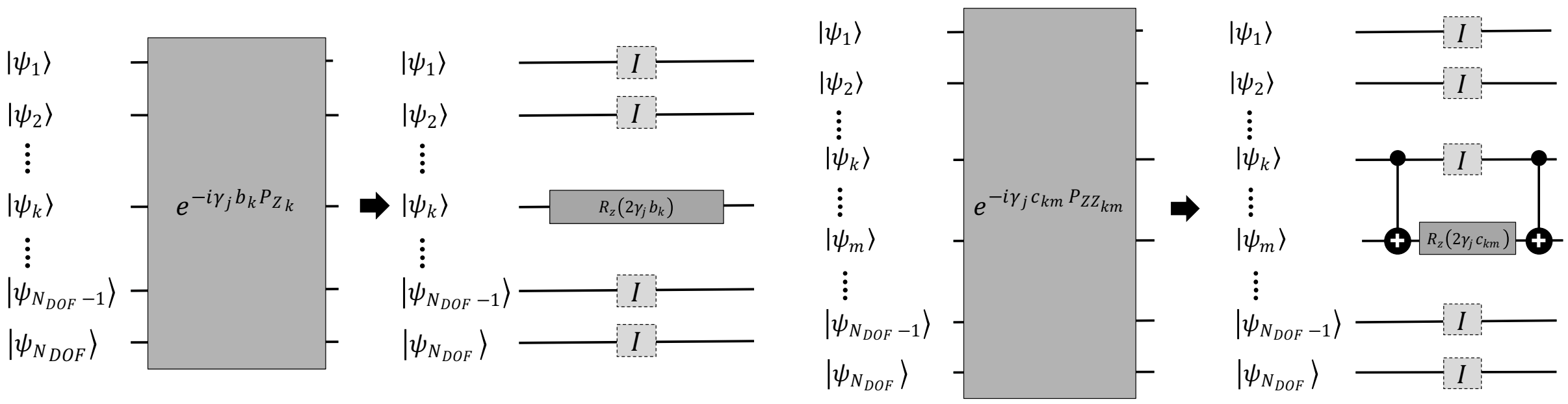

Figure 7: Quantum circuit mapping for Hamiltonian $H_2$. (Left) Single Pauli-Z operators, representing linear variables in the objective function, are mapped into a single Rotation-Z gate. (Right) Double Pauli-Z operators, representing quadratic terms in the objective function, are mapped into a structure composed of two CNOT gates with a single Rotation-Z gate in between.

As a summary, equation (34) represents a gate-based quantum computing approximation of the quantum adiabatic theorem. In this approximation, the quantum evolution from a base Hamiltonian $H_A$ to a cost Hamiltonian $H_B$ is executed by successively applying the unitary operations $U_{H_A}$ and $U_{H_B}$ to a ground energy state of $H_A$. This is the basis of the QAOA used to solve QUBO problems. While in the quantum adiabatic theorem both the parameters $\beta_j$ and $\gamma_j$ directly depend on the function $s(t)$, the original version of the QAOA algorithm does not model this dependence due to the function $s(t)$, and the final evolution time $T$ being unknown. As such, the QAOA performs a classical optimization loop over the full set of $2P$ parameters $\beta_j$ and $\gamma_j$ to optimize the objective function. A clear disadvantage of this approach is the large number of parameters that need to be optimized, adding complexity to the optimization landscape. As a solution, Brandhofer et. al [37] proposed to harness the monotonically increasing nature of $s(t)$ to model $\{\beta_j\}_{j=1}^{P}$ and $\{\gamma_j\}_{j=1}^{P}$ as monotonically decreasing and increasing sets, respectively[5]. For this, $\beta_j$ and $\gamma_j$ are modeled as:

$$\gamma_j = m_\gamma \cdot \frac{2i-1}{2p} \quad for\ i \in \{1, \dots, P\} \tag{47}$$

$$\beta_j = m_\beta \cdot \left(1 - \frac{2i-1}{2p}\right) \quad for\ i \in \{1, \dots, P\} \tag{48}$$

This linear approach has two advantages. First, the optimization can be performed over only the set $\{m_\gamma, m_\beta\}$ independently of the value of $P$, which greatly simplifies the optimization process. Second, the physical meaning of $\beta$ and $\gamma$, given by the quantum adiabatic theorem, is respected. As such, this approach is used in all experiments shown in this paper.

Finally, the parameter $P$, described in section 4.1 as the number of intervals in which the timespan $[0, T]$ is discretized, can be interpreted as a measure of how exact the approximation will be with respect to the original quantum adiabatic algorithm. As a result, it is expected to obtain higher performance solutions (closer to the optimum) as $P$ increases.

The QAOA is a hybrid algorithm in the sense that it still requires a classical counterpart to optimize the parameter set $\{m_\gamma, m_\beta\}$. If the circuit is formed and measured multiple times, a set of feasible solutions will be obtained. These solutions can then be classically evaluated to obtain a metric for the performance obtained by using certain values of $\{m_\gamma, m_\beta\}$. Finally, a classical optimization scheme can be used to tune these parameters to converge towards an optimum. While the final measurement of a quantum circuit will contain a certain degree of stochasticity, as the cost Hamiltonian guides the search of a lower energy state, it is expected that the measurement process will assign larger and larger probabilities to those solutions close to the optimum.

A nice interpretation of the QAOA technique is to understand it as a black box that transform an optimization problem defined over a discrete domain into one that is defined over a continuous set of variables $\{m_\gamma, m_\beta\}$. As such, it opens the door to a wider variety of classical optimization approaches, such as Stochastic Gradient Descent (SGD). A diagram of the quantum circuit for the QAOA algorithm is shown in Figure 8.

[5] Recall that $\beta_j = \Delta t\big(1 - s(j\Delta t)\big)$ and $\gamma_j = \Delta t\ s(j\Delta t)$, and therefore as $j \to P$, $\beta_j$ will decrease and $\gamma_j$ will increase.

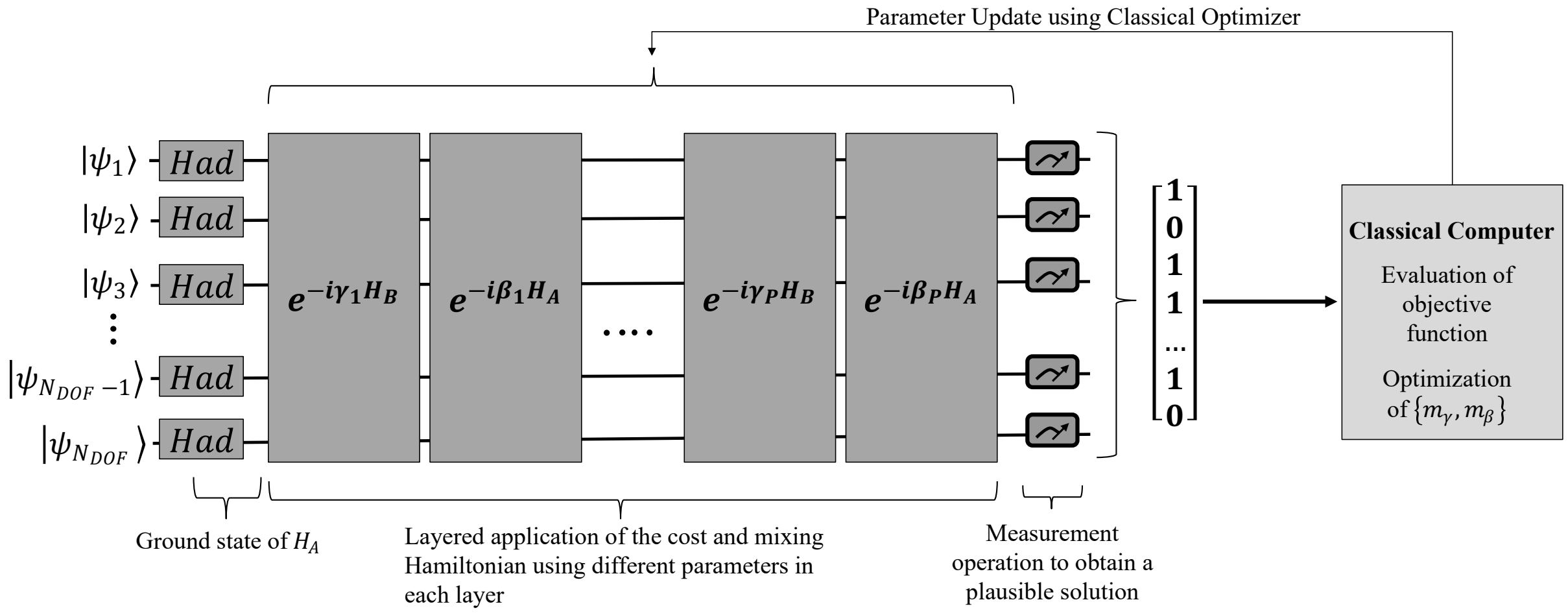

Figure 8: Quantum Approximate Optimization Algorithm framework.

### 4.4 Constrained optimization using QAOA.

The previous sections described how gate-based quantum computing can be utilized to solve quadratic unconstrained binary optimization (QUBO) problems. However, many practical problems require the modeling of constraints on their decision variables. For this paper, the exposition will be limited to linear equality constraints on the number of sensors, as this is the most common type of constraint found in OSP literature. Equation (49) shows a linear equality constrains, whose parameters are contained in the diagonal matrix $Q_3 \in \mathbb{R}^{N_{DOF}} \times \mathbb{R}^{N_{DOF}}$:

$$Q_3\vec{x} = n_s \tag{49}$$

Currently, there are two main approaches to incorporating these types of constraints into the QAOA framework. The first approach is to add the linear constraint as a penalization term of the form $\alpha(Q_3\,\vec{x} - n_s)^2$ into the objective function. The parameter $\alpha$ controls how much a deviation with respect to the constrain is penalized. If finely tuned, the optimization algorithm will naturally avoid unfeasible solutions due to their high cost. A natural advantage of this approach is that the penalization term is also a quadratic function and, therefore, can be naturally incorporated and encoded into a quantum circuit using the techniques reviewed in section 4.2. However, this approach incorporates into the algorithm the complexity of finding the correct value for the hyperparameter $\alpha$. The second approach offers a solution for this problem under the condition that the matrix $Q_3 = I$. Under this consideration, the literature proposes for the Hamiltonian $H_A$, presented in section 4.3, to be changed for a "XY" Hamiltonian [37], $H_{XY}$, shown in equation (50):

$$H_{XY} = \sum_{(k,m)\in G} P_{X_k}P_{X_m} + P_{Y_k}P_{Y_m} \tag{50}$$

where $G$ is a set containing all the pairs of the form $\{(1,2),(1,3),(1,4),\dots(N_{DOF}-1,N_{DOF})\}$, without repeating indices. As expected, the terms in $H_{XY}$ commute with each other and therefore its associated unitary operation can be written as [37]:

$$R_{km}^{XY}(\beta_j) = e^{i\beta_j(P_{X_k}P_{X_m}+P_{Y_k}P_{Y_m})} = \begin{bmatrix} 1 & 0 & 0 & 0 \\ 0 & \cos\left(\frac{\beta_j}{2}\right) & i\sin\left(\frac{\beta_j}{2}\right) & 0 \\ 0 & i\sin\left(\frac{\beta_j}{2}\right) & \cos\left(\frac{\beta_j}{2}\right) & 0 \\ 0 & 0 & 0 & 1 \end{bmatrix} \tag{51}$$

When an XY Hamiltonian is used instead of the Traditional Hamiltonian $H_A$ presented in section 4.3, the initial state needs to change to fulfill the requirement of being a ground state of $H_{XY}$. The new initial state is known as a Dicke state $|D_{n_s}^{N_{DOF}}\rangle$, and its formulation is shown in equation (52), as described in [37]:

$$|D_{n_s}^{N_{DOF}}\rangle = \frac{1}{\sqrt{\binom{N_{DOF}}{n_s}}} \sum_{\substack{k_1,\dots,k_{N_{DOF}}\in\{0,1\} \\ k_1+k_2+\cdots+k_{N_{DOF}}}} |k_1\, k_2 \dots k_{N_{DOF}}\rangle \tag{52}$$

This initial state assigns equal probabilities to all those states that contain exactly $n_s$ entries equal to one. The XY Hamiltonian $H_{XY}$ will then perform a "swapping" operation between two qubits at a time (changing a pair of qubits from 0 to 1 and from 1 to 0), keeping the number of qubits finally measured as 1 constant and therefore exclusively exploring solutions that are feasible under the constrain $I\vec{x} = \sum_{i=1}^{N_{DOF}} x_i = n_s$. This approach to tackling linear constraints has the advantage of not adding the $\alpha$ hyperparameter to the model. However, it is less flexible since it can only deal with constraints of the form shown in equation (49). Additionally, the generation of the Dicke state and the implementation of the XY Hamiltonian $H_{XY}$ require more quantum gates than the Traditional Hamiltonian $H_A$ described in section 4.3.

Both approaches will be tested in the numerical case studies presented in this paper.

# 5 Proposed Framework for OSP using QAOA.

As mentioned in section 2, the QAOA-based framework to find near-optimal sensor configurations in civil structures uses the Modal Strain Energy (MSE) as an objective function [22]. The reason for this is the quadratic form of the MSE function, which makes it ideal for its translation as a QUBO instance. Since the QAOA is a combinatorial optimization solver, the MSE is modified to include binary variables that represent the selection of sensor positions. Equation (53) presents the modified version of the MSE, where the dependency on the set of selected sensor positions $S$ (as shown in equation (3)) is dropped in favor of the binary variables $x_p$ and $x_q$:

$$MSE = \sum_{i=1}^{N_{DOF}} \sum_{j=1}^{N_{DOF}} \sum_{p=1}^{N_{DOF}} \sum_{q=1}^{N_{DOF}} |\phi_{pi} k_{pq} \phi_{qj}|\, x_p x_q \tag{53}$$

As the term $|\phi_{pi} k_{pq} \phi_{qj}|$ is completely determined by the dynamic properties of the structure, equation (53) represents a quadratic objective function dependent on the set of binary variables $\{x_1, \dots, x_{N_{DOF}}\}$.

For this paper, we limit ourselves to only considering equality constraints, i.e., we have a budged of $n_s$ sensors to install in the structure. This constraint is presented in equation (54):

$$\sum_{i=1}^{N_{DOF}} x_i = n_s \tag{54}$$

Consequently, the optimization problem that is going to be solved using the QAOA approach is described in equation (55).

$$\begin{gathered} \max \sum_{i=1}^{N_{DOF}} \sum_{j=1}^{N_{DOF}} \sum_{p=1}^{N_{DOF}} \sum_{q=1}^{N_{DOF}} \left|\phi_{pi} k_{pq} \phi_{qj}\right| x_p x_q \\ s.t. \quad \sum_{i=1}^{N_{DOF}} x_i = n_s \end{gathered} \tag{55}$$

When the equality constraint is introduced using a Traditional Hamiltonian $H_A$, the objective function is modified with the corresponding penalization term, as shown in equation (56).

$$\max \sum_{i=1}^{N_{DOF}} \sum_{j=1}^{N_{DOF}} \sum_{p=1}^{N_{DOF}} \sum_{q=1}^{N_{DOF}} \left|\phi_{pi} k_{pq} \phi_{qj}\right| x_p x_q - \alpha \left( \sum_{i=1}^{N_{DOF}} x_i - n_s \right)^2 \tag{56}$$

The proposed framework to solve equation (55) or equation (56) using the QAOA approach is listed below as an eight-step process:

1. Create a FEM model of the structure to obtain the stiffness and mass matrices.
2. Solve the eigenvalue and eigenvector problem posed by equation (2) to obtain the modal shape matrix $\mathbf{\Phi}$.
3. Use equation (53) to compute the coefficients $\sum_{i=1}^{N_{DOF}} \sum_{j=1}^{N_{DOF}} \left|\phi_{pi} k_{pq} \phi_{qj}\right|$ for each of the terms $x_p x_q$ of the QUBO objective function using the MSE criteria. Recall from section 4 that $x_p x_p = x_p$ since $x_p$ is a binary variable.
4. Apply the relevant constraints to the problem. In particular, the proposed framework considers that the number of sensors that can be placed on the structure is equal to $n_s$. As such, a restriction of the type $\sum_{i=1}^{N_{DOF}} x_i = n_s$ must be introduced via either a penalization term or the use of the XY mixer Hamiltonian, following the processes explained in section 4.4.
5. Using the mapping strategies described in section 4.3, generate the circuit corresponding to $e^{-i\gamma_j H_B}$ and $e^{-i\beta_j H_A}$ or $e^{-i\beta_j H_{XY}}$, depending on the penalization strategy chosen in step 4. Repeat this structure $P$ times ($j \in \{1, \dots, P\}$) and finalize with a measurement operation.
6. The generated quantum circuit is executed and measured $n_e$ times. During each measurement, a binary vector $\vec{x}$ representing a solution for the optimization problem is obtained. Using the objective function, the cost of each of these solution vectors is computed and averaged.
7. A classical optimizer is used in conjunction with the average cost to tune the parameters $\{m_\gamma, m_\beta\}$ towards a local optimum. The quantum circuit is treated as a black box function for this process, i.e., cost $= QAOA(m_\gamma, m_\beta)$.
8. Steps 6 and 7 are repeated until a stop criterion is reached (either a metric of performance, a maximum number of iterations or any other criteria particular to the classical optimizer used).

Once the parameters $\{m_\gamma, m_\beta\}$ have been successfully tuned, a final set of quantum circuit measurement are executed to estimate the discrete probability distribution over the space of feasible solutions. The main objective of this paper is to assess how well this probability distribution can represent the solution of the original COP problem.

Finally, it is important to assess the number of gates required by this proposed approach to map an optimization problem into a gate-based quantum computer as a function of the structure's complexity and desired degree of accuracy. For this analysis, we will assume the worst-case scenario, which is a matrix $\boldsymbol{K}$ and $\boldsymbol{\Phi}$ such that all the pairs $x_p x_q$ have a non-zero coefficient in the objective function. While this is seldom to occur in practice, given that real matrices $\boldsymbol{K}$ and $\boldsymbol{\Phi}$ usually contain zero entries, it allows us to study the upper limit in the number of gates as the size of structure increases. Under the aforementioned conditions, and assuming that the total number of degrees of freedom is $N_{DOF}$, equation (53) indicates that the maximum number of unique and independent pairs $x_p x_q$ with $p \neq q$ in the objective function is given by $(N_{DOF}^2 - N_{DOF})/2$ while $N_{DOF}$ terms are of the form $x_p x_p$. Therefore, the objective function will have $(N_{DOF}^2 - N_{DOF})/2$ quadratic terms and $N_{DOF}$ linear terms[6]. The extra terms added to the objective function due to an equality linear constraint using the penalization term technique can be factorized into these terms, and thus do not produce extra gates in the quantum circuit. As discussed in section 4.2, each quadratic term in $H_B$ will require three gates[7], while each linear term will only require one. As such, the size of the unitary operation $U_{H_B}$ is at most $3(N_{DOF}^2 - N_{DOF})/2 + N_{DOF} = 3N_{DOF}^2/2 - N_{DOF}/2$ gates. On the other hand, the number of gates generated by to $H_A$ will depend linearly on the number of variables. Therefore, the size of the overall quantum circuit is controlled by the cost Hamiltonian encoding and is quadratic on $N_{DOF}$. Figure 9 shows how the number of quantum gates scale with the number of degrees of freedom in the structure for different values of $P$.

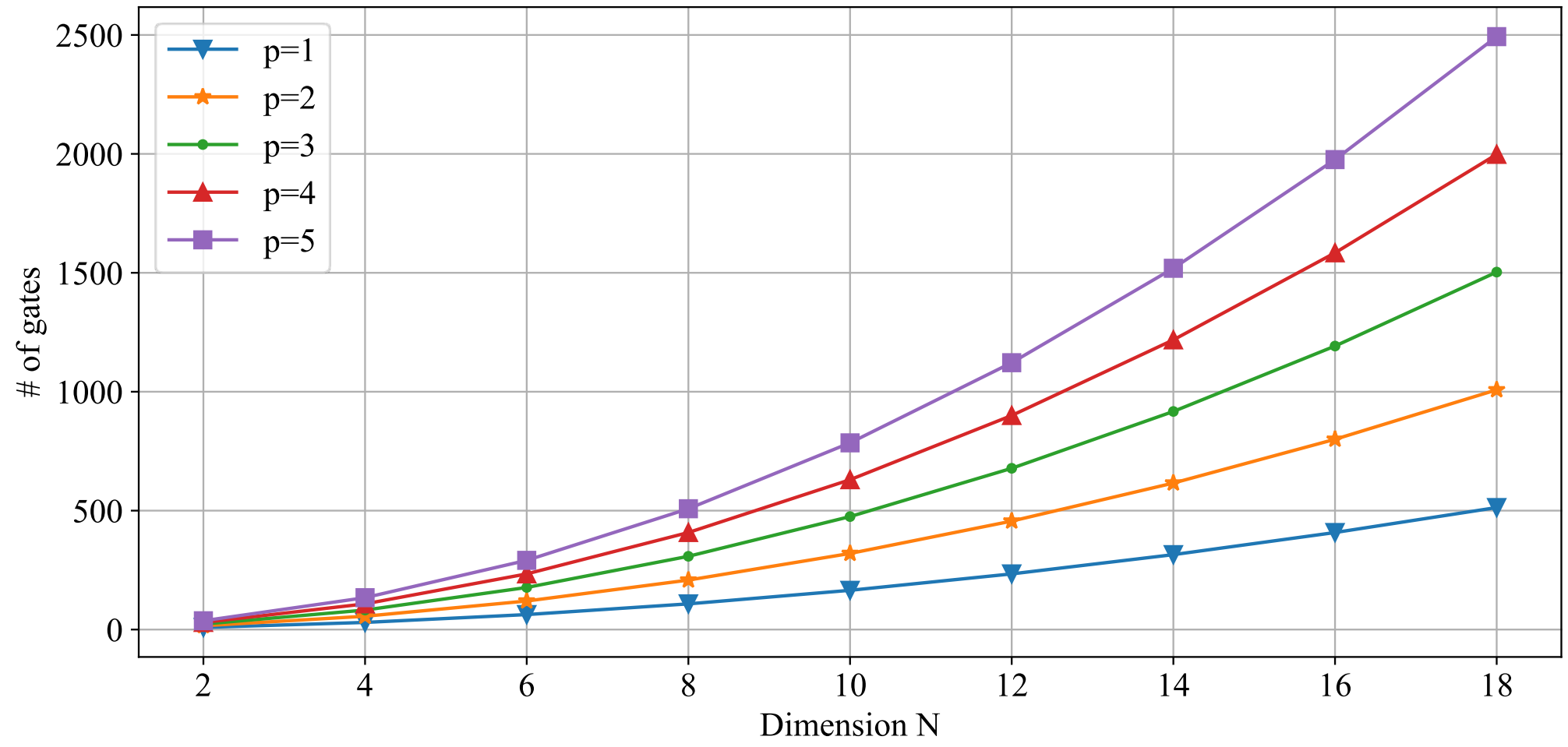

Figure 9: Number of quantum gates required to map the OSP problem for a structure of $N_{DOF}$ degrees of freedom, using multiple levels of repetitions in the QAOA algorithm.

# 6 Experimental Validation

This section presents results regarding the application of the proposed framework to two case studies of varying size. However, before delving into the case studies description and the results obtained by the QAOA approach, it is necessary to list a set of experimental considerations taken into account for the execution of the quantum circuits.

---

[6] Since $x_p$ represent binary variables, terms of the form $x_p x_p$ are equivalent to $x_p$.

[7] Remember that each quadratic term is mapped using two CNOT gates and one Rotation-Z gate.

## 6.1 Experimental considerations

### 6.1.1 Determination of the penalization parameter $\alpha$

Experimentally, we found that the QAOA performed better when the objective values are in the range comprehended between ~1 and ~10, approximately. For this purpose, we divided each one of the QUBO terms $\sum_{i=1}^{N_{DOF}} \sum_{j=1}^{N_{DOF}} |\phi_{pi} k_{pq} \phi_{qj}|$ in the objective function by a constant corresponding to the total modal strain energy contained in the structure. By doing this, we are effectively normalizing the problem to obtain values in the aforementioned range. This modification of the objective function has the following advantages: (i) it preserves the ranking of solutions, which is fundamental since we do not want to alter the optimization problem, just to normalize it, (ii) it only requires to know the value of the total modal strain energy contained in the structure (assuming sensors at all degrees of freedom), which is easily computable by a simple matrix multiplication operation concerning the stiffness and modal matrices and does not require any type of optimization procedure, and (iii) since the problem is already normalized, we can define $\alpha = 1$ and still have a well-scaled penalization term, in which all the feasible solutions have higher objective values than the non-feasible ones. With this approach, we avoid the cumbersome process of defining the parameter $\alpha$, which usually requires extensive testing and is problem dependent.

### 6.1.2 Performance measurement

For reporting purposes, the raw value of the MSE achieved by the solutions obtained with the QAOA approach does not give a clear indication of how good or bad the solutions are. To overcome this challenge, we will follow the approach described by Brandhofer et al. [37]and defined a *normalized performance* metric, $r$, shown in equation (57) for a given candidate solution $\vec{x}_i \in \{\vec{x}_i\}_{i=1}^{2^{N_{DOF}}}$.

$$r(\vec{x}_i) = \begin{cases} \dfrac{MSE(\vec{x}_i) - MSE_{min}}{MSE_{max} - MSE_{min}}, & if \displaystyle\sum_{j=1}^{N_{DOF}} x_{ij} = n_s \\ 0, & if \displaystyle\sum_{j=1}^{N_{DOF}} x_{ij} \neq n_s \end{cases} \quad (57)$$

where $x_{ij}$ is the $j$-th component of the candidate solution $\vec{x}_i$, and $MSE_{max}$ and $MSE_{min}$ are the highest and lowest modal strain energy that can be obtained using a feasible configuration, respectively. In equation (57) a performance of $r = 0$ is automatically assigned to all infeasible solutions. On the other hand, the case $r = 1$ will corresponds to the optimal solution. Note that in a practical OSP problem, this normalization performance metric would be impossible to compute, since it requires knowledge of global maximum and minimum solution. However, the case studies presented in this paper are designed to be small enough that an extensive-search solution is possible in a couple of minutes, hence allowing for the normalized performance to be computed.

As explained in section 3.4, the output of QAOA's quantum circuit is a quantum state that can be interpreted as a discrete probability distribution over the space of feasible and non-feasible solutions, $p_{QAOA}(\vec{x}_i)$. Using equation (57), we can give a performance metric for how well this distribution assigns probability to optimal solutions by computing:

$$r(p_{QAOA}) = \sum_{i=1}^{2^{N_{DOF}}} p_{QAOA}(\vec{x}_i) \cdot r(\vec{x}_i) \quad (58)$$

where a weighted normalized performance is computed using as weights the probability values assigns to each possible feasible and non-feasible solution.

6.1.3 Computational and Optimization considerations

Due to the general unavailability of quantum hardware at the moment of writing this paper, all the results presented in this section were obtained using a quantum simulator program, which is a piece of code that can be executed in a traditional computer to simulate the behavior of a quantum computer. This places important limitations on the size of structures that can be analyzed. In particular, since each qubit in the quantum computer represents a degree of freedom in the structure, we are limited to analyze structures of ~20 degrees of freedom. The reason for this limitation is the exponential nature of quantum states with respect to the number of qubits. As mentioned in section 3.3, unitary matrices reach a size of $2^n \times 2^n$, where $n$ is the number of qubits in the system. As a consequence, the analysis of a structure with $20$ candidate locations for sensors will require the execution of operation concerning matrices with $2^{20} \times 2^{20} = 1.099 \times 10^{12}$ elements, which is already at the limit of what a desktop computer can handle comfortably for experimentation purposes.

The quantum simulator used was PennyLane 0.33.1 [38], executed as a Python 3.10.8 library in a computer with 128 GB of RAM, and an AMD Ryzen Threadripper PRO 3955WX 16-Cores processor. For the classical optimization loop, we have used the ADAM optimizer [39] to found optimal values for $\{m_\gamma, m_\beta\}$. The optimization termination criteria were the same for all experiments and correspond to finalize the optimization cycle if 500 iterations have been achieved or if the objective function has not been improved above a certain tolerance in three consecutive iterations. As a tolerance for improvement, we used $\epsilon = 10^{-4}$ .

## 6.2 Case studies description

We have chosen to test the QAOA proposed methodology on the same system at different scales to better compare how does the algorithm performs as the size of the structure increases. Consequently, the approach proposed in section 5 will be experimentally tested on Warren truss bridges of five (WT5) and 11 nodes (WT11), depicted in Figure 10(a) and Figure 10(b), respectively. The non-supported degrees of freedom are noted in each figure. All truss elements in both structures have the following properties: cross-sectional area of $A = 1.43 \times 10^{-2} m^2$, Young's modulus $E = 200\ GPa$, density $\rho = 7850 \frac{kg}{m^3}$ and length $L = 2m$. The stiffness matrix and mass matrices for both structures are computed using a finite element approach, employing truss elements and equivalent lump masses at the nodes. After the supported degrees of freedom are released from the structures, the candidate locations to place sensors (corresponding to the non-supported degrees of freedom) are reduced to six for WT5 and 18 for WT11. Additionally, the modal mass participation factors assuming a uniform excitation vector in the horizontal and vertical directions were computed to select a set of target modes for both structures. For the results used in this paper, we selected a set of the most significant modes such that they contributed 95% of the total effective mass. Following this approach, the target modes selected for WT5 are mode 1, mode 2 and mode 5, while for WT11 mode 1 through mode 4.

As mentioned before, these structures are small enough that their optimal sensor configuration can be computed using an exhaustive search approach. Table 4 lists the total number of possible sensor configurations for both structures.

Table 5 lists the 5 feasible candidate sensor configurations with the highest normalized performance for each structure, following equation (57).

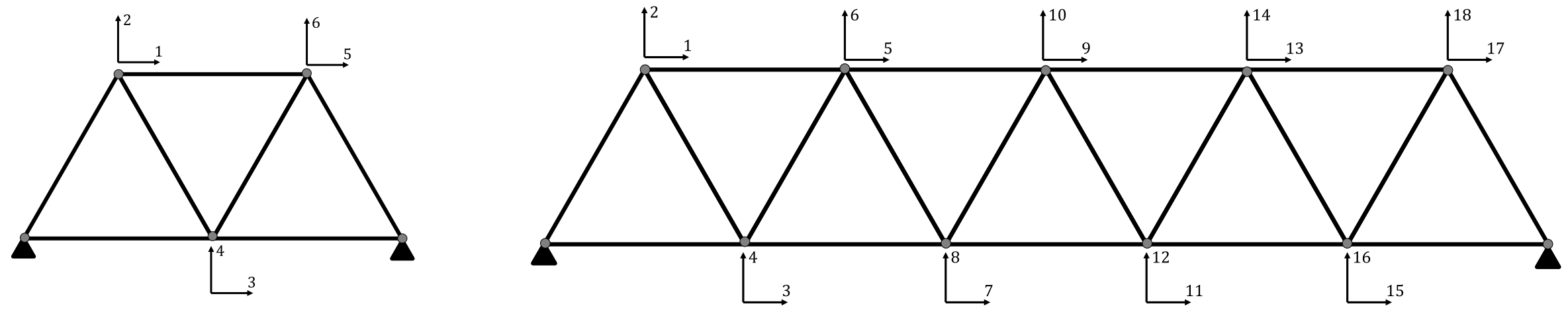

(a) Warren truss bridge with five nodes (WT5)

(b) Warren truss bridge with 11 nodes (WT11)

Figure 10: Structures considered in the case studies.

Table 4: Total number of sensor configurations and number of feasible sensor configurations for both case studies presented in this paper.

| | **WT5** | **WT11** |
|---|---|---|
| Total number of sensor configurations | $2^{N_{DOF}} = 2^6 = 64$ | $2^{N_{DOF}} = 2^{18} = 262{,}144$ |
| Number of feasible sensor configurations | $\binom{N_{DOF}}{n_s} = \binom{6}{2} = 15$ | $\binom{N_{DOF}}{n_s} = \binom{18}{4} = 3060$ |

Table 5: top 5 optimal solutions obtained by performing an exhaustive search approach for both case studies.

| | **WT5** | | **WT11** | |
|---|---|---|---|---|
| **Solution Ranking** | **Normalized Performance** | **Sensor locations [DOFs]** | **Normalized Performance** | **Sensor locations [DOFs]** |
| 1 | 1.0 | 1, 5 | 1.0 | 4, 6, 14, 16 |
| 2 | 0.994 | 2, 3 | 0.968 | 10, 12, 14, 16 |
| 3 | 0.994 | 3, 6 | 0.968 | 4, 6, 8, 10 |
| 4 | 0.793 | 3, 5 | 0.963 | 2, 4, 6, 8 |
| 5 | 0.793 | 1, 3 | 0.963 | 12, 14, 16, 18 |

### 6.3 Experimental results

As mentioned in section 4, the QAOA approach can be understood as a black-box function that changes the original discrete optimization problem into a continuous optimization problem dependent on the parameter set $\{m_\gamma, m_\beta\}$ through equations (47) and (48). We start this section by exploring what are the characteristics of the normalized performance function produced by the QAOA algorithm. For this, Figure 11 (a) and (b) show the normalized performance landscape obtained in both case studies, respectively. To produce these plots, we have varied the number of circuit repetitions (rows) and the Hamiltonian used as a mixer (columns). The color bar range indicates normalized performances, which varies from 0 to 1. In the title of each plot, we have noted the maximum normalized performance achieved in each particular case. It is important to note that these are not optimization results, but the landscapes of the normalized performance function obtained by just varying the parameters $m_\beta$ and $m_\gamma$ from 0 to $2\pi$.

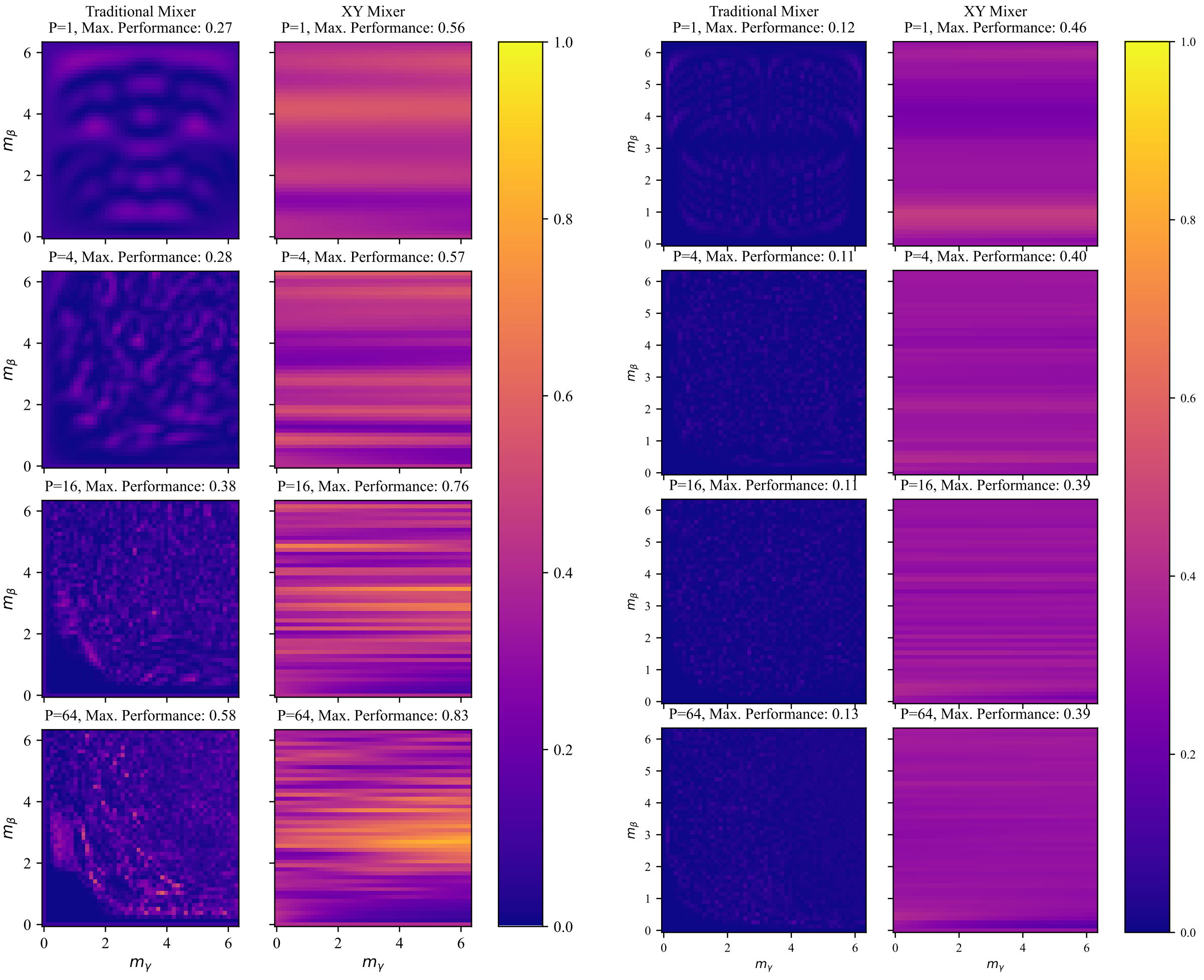

(a) Warren truss bridge of five nodes (WT5) (b) Warren truss bridge of 11 nodes (WT11)

Figure 11: Optimization landscape for the (a) WT5 case study and (b) WT11 case study. In each case, the left column presents results obtained using a Traditional Hamiltonian mixer ($H_A$), while the second column presents results obtained using a XY Hamiltonian mixer $H_{XY}$. Rows indicates different values for the number of circuit repetitions, $P$. Finally, the title of each subplot reveals the maximum normalized performance achieved in each case.

In Figure 11(a), which corresponds to the WT5 case study, several notable observations emerge. Firstly, it is evident that higher values of parameter $P$, corresponding to the number of QAOA circuit repetitions, strongly correlate with the highest normalized performance achieved. Secondly, when considering an equivalent number of circuit repetitions, the XY mixer Hamiltonian consistently outperforms the Traditional Hamiltonian in terms of normalized performance. This outcome aligns with our initial expectations, as the XY mixer Hamiltonian inherently incorporates the equality constrain related to the number of sensors to be placed in the structure, thereby constraining the search space accessible to the QAOA. On the other hand, the Traditional mixer Hamiltonian uses a penalization approach to enforce this constraint, which while effective, it does not reduce the search space available to QAOA.

Additionally, while for the most part the Traditional Hamiltonian landscapes lack a discernible structure, the XY Hamiltonian exhibits a distinctive horizontal "stripped" pattern. This behavior suggests that the concentration of higher performance levels occurs at specific values of $m_\beta$ with a relatively lower influence of the parameter $m_\gamma$. This observation provides valuable insights into the functionality of the XY Hamiltonian. It's important to recall that $m_\beta$ is associated with the mixer Hamiltonian, dictating the states explored during the quantum optimization process. Conversely, $m_\gamma$ is linked to the cost Hamiltonian, determining the states which will be assigned a higher probability based on their superior normalized performance. Considering these roles, it is logical that for a smaller structure—where there are only 15 feasible candidate locations that the XY Hamiltonian will explore— the selection of which states to explore domain the overall normalized performance evaluation.

In the context of the case study concerning WT11, as depicted in Figure 11(b), several key observations emerge. First and foremost, unlike the WT5 case study, there is no evident improvement in performance with an increase in the number of circuit repetitions, $P$. This difference in behavior hints at a possible shortcoming with the linearized approach to define the parameter set $\{\beta_j, \gamma_j\}_{j=1}^{P}$ from the set $\{m_\beta, m_\gamma\}$, where for larger structures it may be necessary to explore non-linear dependances. On a different note, the XY Hamiltonian continues to be the preferable approach in terms of normalized performance when compared against Traditional Hamiltonian. The observed difference in normalized performance between both Hamiltonians enforces our intuition that reducing the available solution space by structurally integrating the existing constraints into the QAOA circuit is conducive to better overall performance.

In general terms, the landscapes produced by the QAOA in the WT11 can be classified as “barren plateaus” [40] given the lack of clear maxima or minima locations. This uniformity hints at an increased level of difficulty in pinpointing global optimal points, probably due to the use of a linear approach to compute the parameter set $\{\beta_j, \gamma_j\}_{j=1}^{P}$ in a larger and more complex case study. Despite this difficulty, a subtle stripped pattern can still be observed for the $H_{XY}$ mixer Hamiltonian, reinforcing our intuition with respect to the different roles that $m_\beta$ and $m_\gamma$ play in the optimization of the QAOA circuit. In stark contrast, the Traditional Hamiltonian loses all identifiable patterns, explaining the poorer results obtained.

Now we turn our attention to the results obtained when the parameters $\{m_\beta, m_\gamma\}$ are optimized. For this, we initialized the ADAM optimizer with the best points identified in the respective heatmap plots. These favorable initial conditions are employed to evaluate the potential effectiveness of the QAOA approach. The optimization cycle uses the considerations outlined in section 6.1.3. Through our experimentation with the ADAM optimizer, we've observed that an increase in the number of optimization cycles typically does not correspond to a subsequent increase in final performance. This behavior can be attributed to the optimization process often converging to a local optimum point within approximately 100 iterations. Subsequent iterations tend to refine the parameters towards points that provide only marginal improvements to the solution.

To ensure robustness, each optimization run was repeated five times, and the standard deviation was represented as error bars to provide insights into the stability of the outcomes. Figure 12 illustrates the normalized performance attained after this optimization process for both case studies. This approach not only allows for a detailed examination of the optimized results but also provides a measure of the variability in performance achieved through repeated optimization runs.

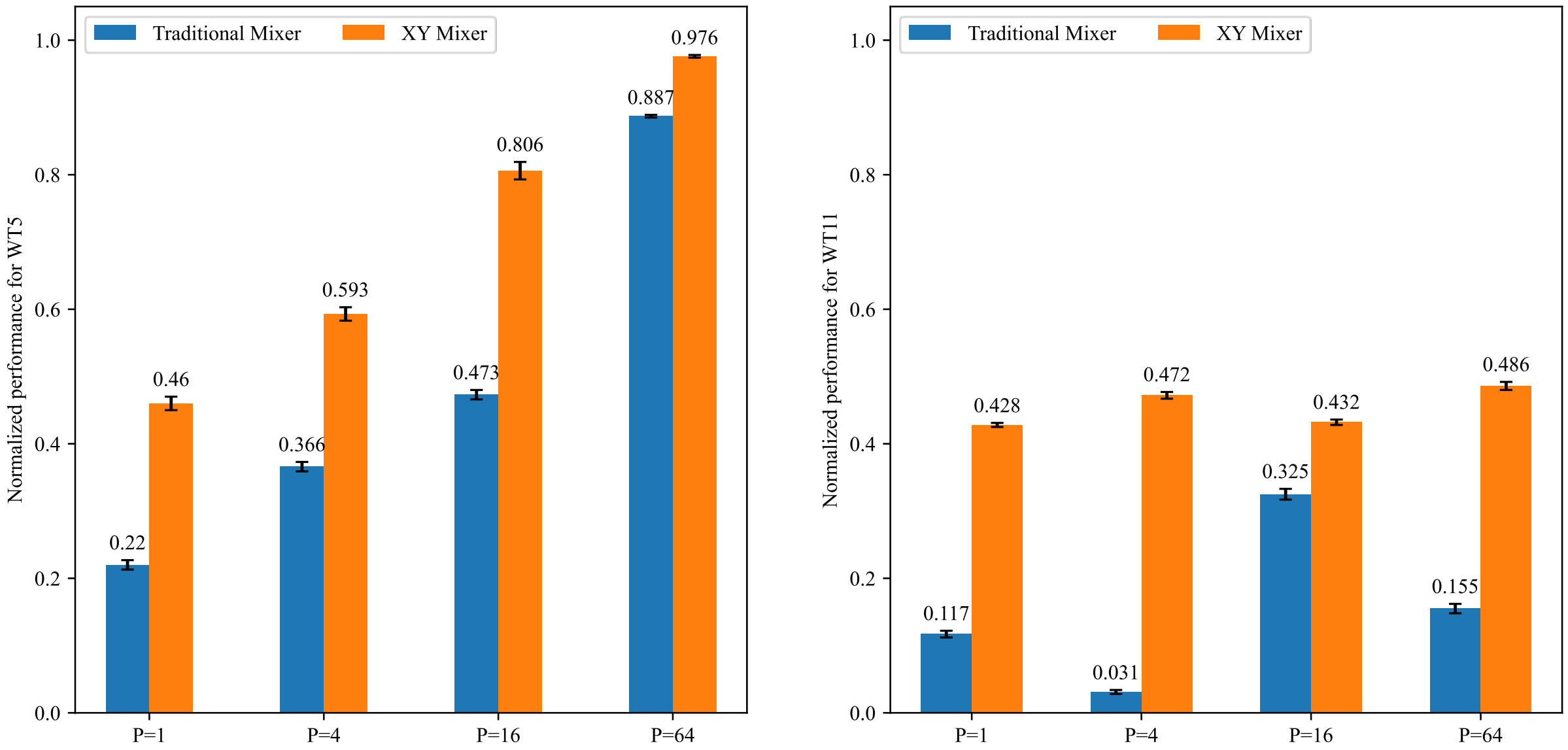

(a) Normalized performance for WT5 (b) Normalized performance for WT11

Figure 12: Normalized performance achieved for both case studies using the Traditional mixer Hamiltonian ($H_A$, in blue) and the XY mixer Hamiltonian ($H_{XY}$, in orange) for varied number of circuit repetitions, $P$.

In the context of the WT5 case study, as depicted in Figure 12 (a), the normalized performance exhibits a clear increase with a higher number of circuit repetitions. This outcome aligns with expectations stemming from the utilization of the Quantum Adiabatic Theorem in the QAOA derivation, where the parameter $P$ represents the granularity for the discretization of the total adiabatic time. Analyzing the choice of Hamiltonian for the mixing operation, it is evident that the XY mixer consistently outperforms the Traditional Hamiltonian across all cases. This result is in accordance with our expectations, as the XY Hamiltonian's capacity to constrain the search space to feasible solutions should inherently result in significant advantages in terms of performance. Notably, for the highest number of circuit repetitions tested ($P = 64$) mixer Hamiltonian achieve a very high normalized performance, indicating that the QAOA is capable of identifying optimal sensor configurations for the WT5 case study structure.

Examining the results for the WT11 case study, illustrated in Figure 12 (b), we can observe that the Traditional mixer Hamiltonian exhibits an irregular performance as the value of $P$ increases. The XY mixer Hamiltonian exhibits a similar, though more consistent, trend. The normalized performance remains largely constant, with a notable increase observed only in the most extreme case of $P = 64$. The fact that performance does not clearly improve as the values of $P$ increases for the WT11 case study can be interpreted as evidence that, for larger structures, the linear process used to define the parameter set $\{\beta_j, \gamma_j\}_{j=1}^{P}$ from $\{m_\beta, m_\gamma\}$ may not be sufficiently flexible. It is essential to consider that, despite the lower final overall performances in comparison to the smaller case study WT5, the larger structure of WT11 introduces a significantly expanded search space. Matter of fact, as shown in Table 4, the size of the feasible space of WT11 is roughly 200 times larger than the space of WT5. Consequently, normalized performances ranging between 40%-50% still imply a noteworthy capacity of the algorithm to produce high quality solutions.

A significant drawback of the proposed quantum-based approach is highlighted by the capability of QAOA, particularly when utilizing the Traditional Hamiltonian ($H_A$), to yield infeasible solutions with a non-null probability. However, as the number of circuit repetitions $P$ increases and the algorithm refines its performance, we should expect, at least in theory, a diminish in the probability of sampling a non-feasible solution. Ideally, the probability of sampling an infeasible solution should tend towards zero.

Figure 13 provides a visual representation of the probability of sampling an infeasible solution for both case studies, exclusively focusing on the Traditional mixer Hamiltonian case, since when the XY mixer Hamiltonian is used, this probability is by definition zero, as non-feasible states are not explored.

The analysis of Figure 13 shows that, for the WT5 case study, the probability of sampling an infeasible solution effectively tends toward zero as the number of circuit repetitions increases. Matter of fact, for the case $P = 64$, case study WT5 reaches a probability of sampling infeasible solutions of less than 1%. However, for the WT11 case study, while there is an overall decreasing trend, this is irregular, with a peak in $P = 4$. This result holds considerable significance, since it underscores the fact that for larger case studies, a linear approach for defining the QAOA parameter set may not be a conducive to improved performance.

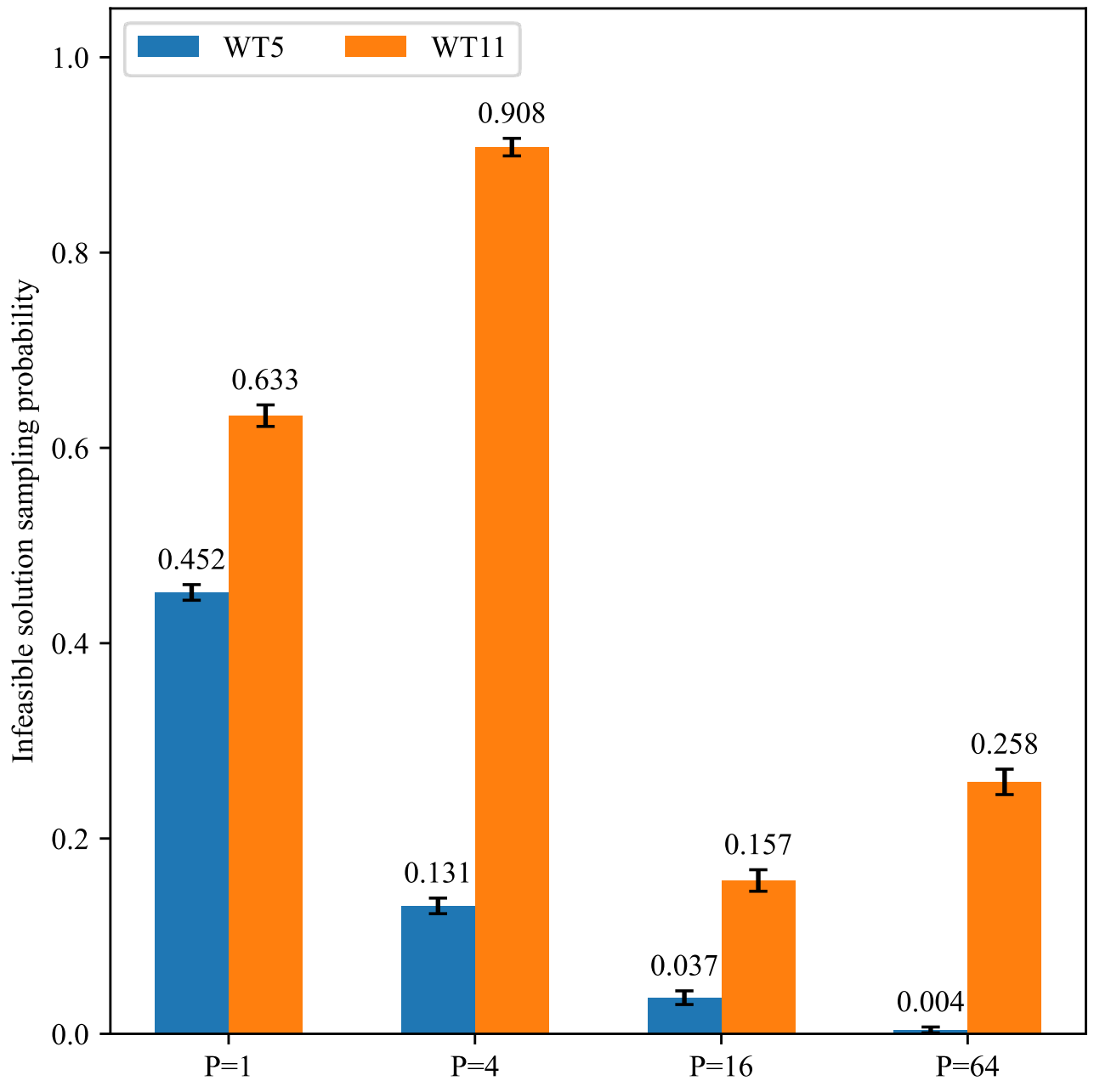

Figure 13: Probability of obtaining infeasible solutions for both case studies using the Traditional mixer Hamiltonian $H_A$. As a reminder, infeasible configurations are those that contain a number of sensors different than two for WT5, or different than 4 for WT11.

Finally, we are interested in the probability of sampling the optimal solution for each particular case study. But before delving into the results, it's crucial to emphasize that a low probability of sampling the optimal solution does not necessarily imply poor model performance. To illustrate this point, consider a hypothetical scenario where a QAOA circuit can sample the optimal solution with only a 2% probability. While this might initially seem inadequate for certain optimization purposes, a closer examination reveals its practical efficacy. In a real-world scenario, this quantum circuit could retrieve the optimal solution, on average, after just 50 executions. Given the computational efficiency of individual circuit executions and the rapid evaluation of the MSE objective function, we can envision a situation where we execute the quantum circuit 500 times, retrieving 500 candidate vectors, and retain the highest-performing solution. In this context, we would have a high degree of confidence that the highest performing solution is in fact, the optimal solution. Consequently, it is not imperative for a model to provide an extremely high probability of sampling the optimal solution to be deemed suitable. What matters more is the model's ability to sample the optimal solution with a non-negligible probability, which is not trivial when the size of the feasible space may contain an enormous number of elements.

Examining Figure 14 for the WT5 case study, we can observe that the probability of sampling the optimal solution increases with the number of circuit repetitions, peaking at 42.7% for the XY Hamiltonian for the case $P = 16$.

However, for $P = 64$, an unexpected decrease in performance occurs for both mixing Hamiltonians, deviating from the anticipated improvement predicted by the QAOA theory. One potential explanation for this unexpected trend lies again in the linear dependence of QAOA parameters $\{\beta_j, \gamma_j\}_{j=1}^{P}$ on the optimization parameter set $\{m_\beta, m_\gamma\}$. It is possible that at higher numbers of circuit repetitions, more flexible dependencies are required to continue improving the results, even for relatively small case studies.

Nevertheless, despite the fluctuations, it is noteworthy that for all cases except $P = 1$, the probability of sampling the optimal solution remains non-negligible. This clearly indicates the potential of QAOA as an algorithm for finding optimal solutions, especially when multiple circuit repetitions are employed.

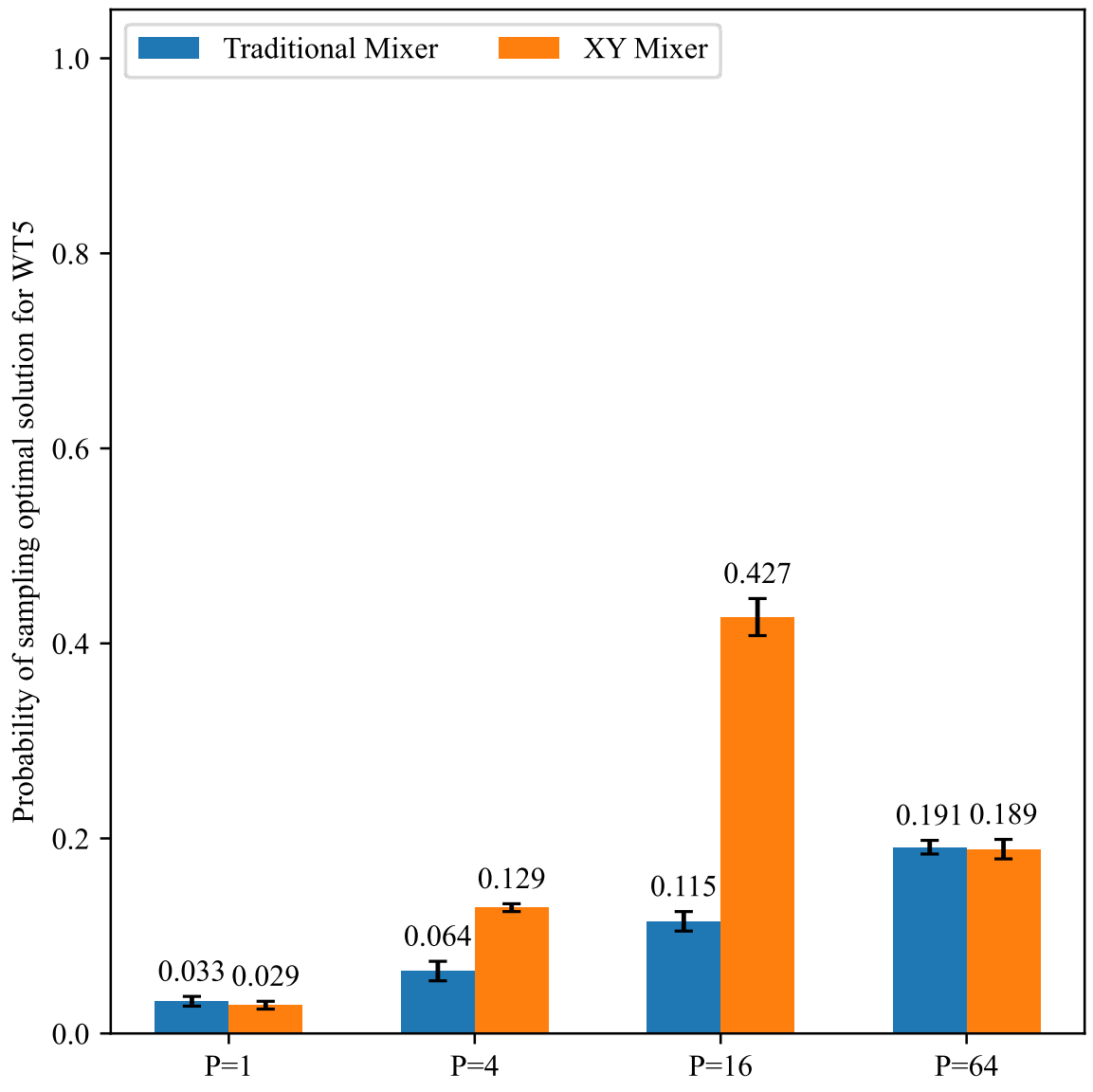

Figure 14: Probability of sampling the optimal solution for the first case study, WT5.

The bar-plot corresponding to the second case study, WT11, was excluded from the previous analysis due to a null probability of sampling the optimal solution for all numbers of circuit repetitions and both mixing Hamiltonians. However, achieving exactly the optimal solution is an extremely uncommon property among heuristic and meta-heuristic algorithms. Most of them are design to return a solution that is close enough to the optimum. To explore the effectiveness of the solutions found by the QAOA approach in the more complex case study, WT11, we relax the condition from requiring the algorithm to sample the optimal solution to only requiring it to sample one of the top 10% best feasible solutions. The results regarding this relaxation are presented in Figure 15. This adjusted analysis provides a more nuanced understanding of the algorithm's performance, allowing for an exploration of its ability to generate solutions that, while not necessarily optimal, rank among the top-performing solutions within the feasible solution space.

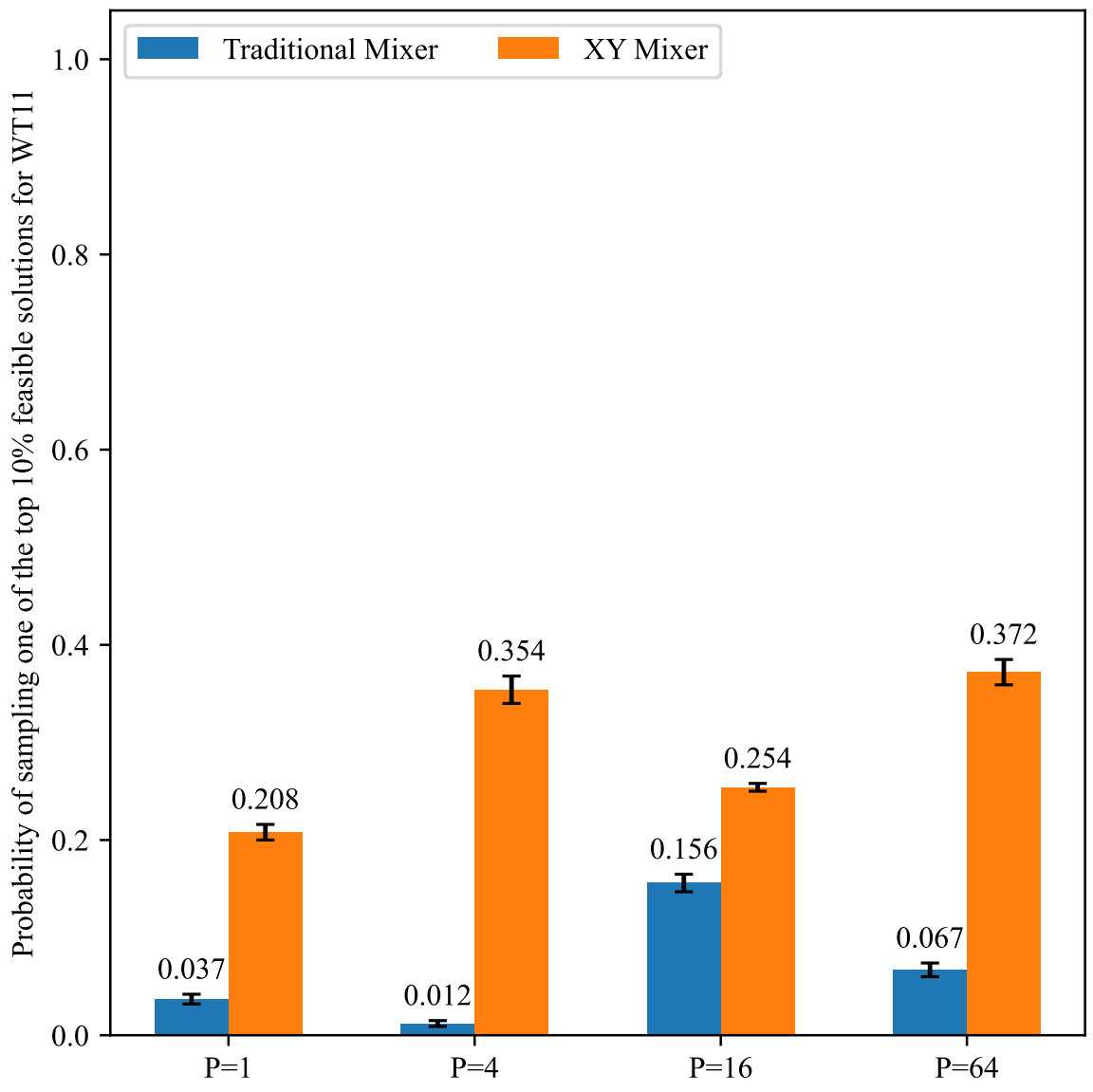

Figure 15: Probability of sampling the top 10% best feasible solutions for the second case study, WT11.

The results from Figure 15 reveal that the probability of sampling one of the top 10% feasible solutions is consistently higher for the XY mixer Hamiltonian, an anticipated result, given its ability to reduce the search space to only solutions that are feasible. The maximum probability achieved in this relaxed condition is 37.2% (for the case $P = 64$ with the XY mixing Hamiltonian). This result demonstrates the QAOA's capability to discover top-performing solutions, even for a relatively complex structure like the WT11 case study.

To conclude this analysis, we would like to draw attention to the results obtained with the traditional Hamiltonian, $H_A$. For the best-performing case, $P = 16$, the probability of sampling a top 10% performing solution is approximately 15%. This raises concerns about whether this Hamiltonian is indeed finding high-performing solutions or merely getting trapped in a local minimum that assign an almost uniform probability to all feasible solutions, with just a modest higher probability weight to near-optimal solutions. In other words, whether the optimization cycle is primarily employed to avoid sampling non-feasible solutions rather than to identify truly superior solutions. To explore this assumption further, Figure 16 (a) and (b) show the results corresponding to the cases in which we want to sample the top 20% and top 50% feasible solutions, respectively.

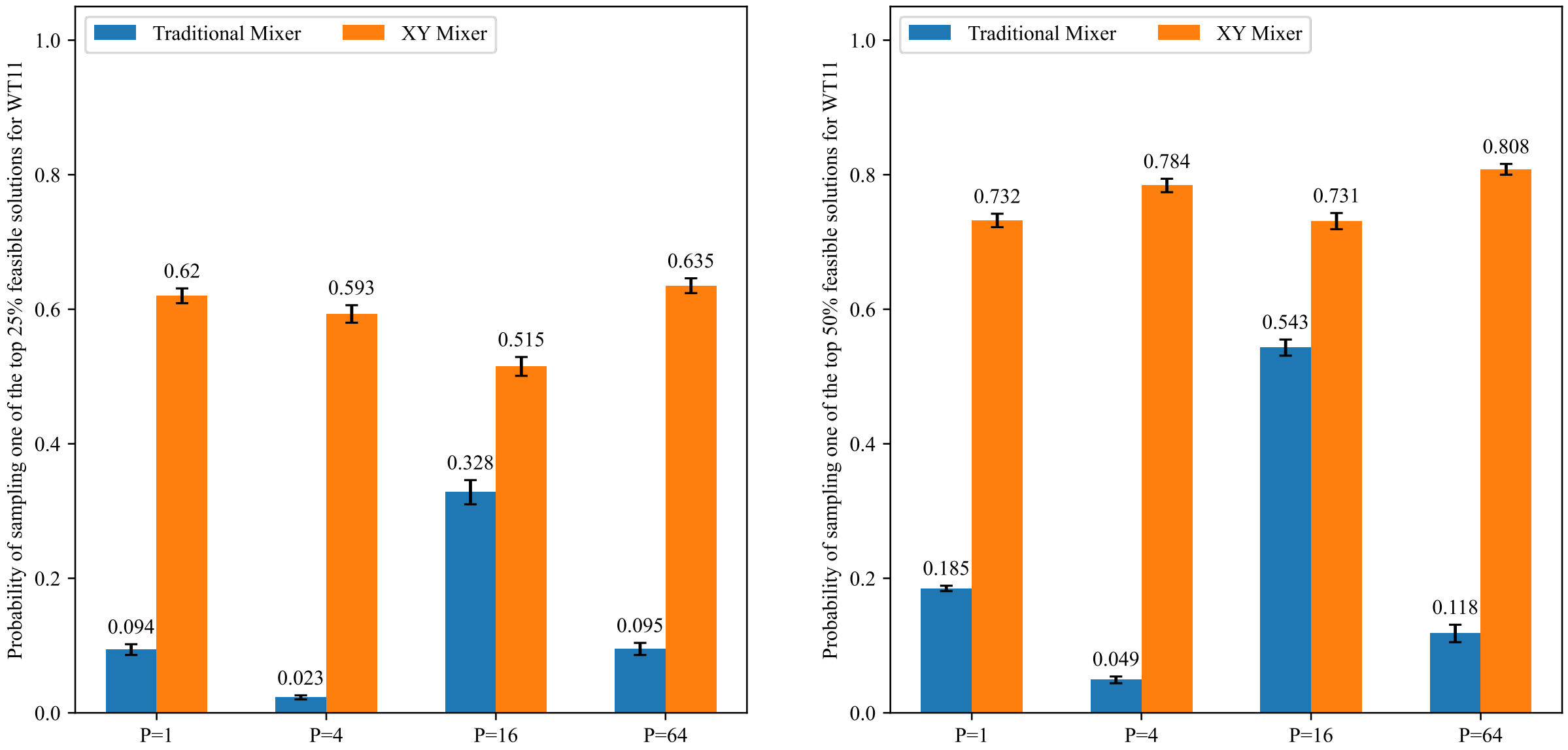

Figure 16: Probability of sampling (a) the top 25% and (b) top 50% best feasible solutions for the second case study, WT11.

As suspected, the observed pattern for the Traditional mixer ($H_A$) persists: the best result achieved are relatively close to the thresholds of 25% and 50% in each case. In contrast, for the XY mixer, the probability of sampling the top 25% or top 50% consistently surpasses these thresholds, for all values of $P$. This pattern suggests that the model utilizing the XY mixer is proficient at assigning a higher probability to solutions that are closer to the optimum, rather than generating a rather uniform probability distribution over the feasible solution space. This discovery holds significant implications, indicating the incapability, at least within the tested levels of circuit repetitions, of the traditional mixer to consistently find optimal solutions. It underscores the importance of incorporating the optimization constraints in a structured manner through the mixer Hamiltonian, as exemplified by the XY mixer.

# 7 Concluding Remarks, Challenges and Future Opportunities.

This paper presented a comprehensive introduction to gate-based quantum computing and quantum-based combinatorial optimization within a structural health monitoring context. For this, a novel framework that uses the Quantum Approximate Optimization Algorithm (QAOA) to solve the optimal sensor placement (OSP) task was proposed to the reader. The framework tackles the OSP task by converting the Mean Strain Energy (MSE) metric into a Quadratic Unconstrained Binary Optimization (QUBO) problem that can be easily encoded into the Hamiltonian of a quantum system and then minimized using the quantum adiabatic theorem. The proposed methodology was applied to numerical models of Warren truss bridges of varying size. The optimization landscape produced by the QAOA was analyzed by performing a search process over the parameter space. The landscape results indicated that the proposed approach is able to generate high quality solutions to the OSP combinatorial problem, especially when the accuracy of the QAOA technique is increased by augmented the size of the associated quantum circuit. Results related to the optimization procedure reveal a similar trend: near-optimal solutions can be discovered by the proposed approach, especially when the optimization constraints relationships are included into the quantum circuit via the mixer Hamiltonian. This is evidenced by the generalized higher results obtained by this approach when compared against the Traditional Hamiltonian that relies on a penalization approach for constraint incorporation.

The results shown in this paper mark a first step towards the study and exploration of current state of the art quantum-based optimization techniques in structural engineering applications. Additionally, the results obtained in this paper highlight challenges that quantum optimization approaches need to overcome to become a suitable alternative to current classical techniques for the OSP problem. In what follows, we describe these challenges and outline possible paths of research to addressed them.

First, the results shows that the QAOA obtains high performant solutions when using a high number of circuit repetitions, especially for the smallest case study. However, more circuit repetitions also imply an increase in the number of gates required to execute the quantum circuit, which imposes limitations in the practical realization of the QAOA approach. For this paper, we avoid any inconveniences related to measurement noise by using a quantum computer simulator. Nevertheless, when using real quantum hardware, experimental results report a maximum of $P \sim 5$ before the quantum state is completely overcome by noise. While an algorithmic solution to noise measurement seems like a less promising path to follow, this issue could be minimized by exploring different optimization strategies that require a smaller number of circuit repetitions to achieve high performing results.

In a similar context as the first challenge, a high number of circuit repetitions also implies the use of a higher number of parameters $\{\beta_i\}_{i=1}^{P}$ and $\{\gamma_i\}_{i=1}^{P}$, which make the optimization process more challenging in practice. For this, the exploitation of physical restrains of the transition function $s(t)$, particularly its monotonically increasing nature, presents a potential pathway to increase the number of circuit repetitions while simultaneously maintaining the optimization problem tractable with a relatively low number of parameters. A linear parameterization strategy was explored in this paper to this very end. However, we observed a clear shortcoming to this technique, especially in the results regarding the largest case study, denoted as WT11. For this, it results crucial for the implementation of the QAOA approach in structures of practical relevance the exploration of different transition functions, ideally non-linear ones, in order to increase the flexibility of the algorithm.

As a third challenge, we can identify the incorporation of constraints from the original optimization problem into the quantum circuit. The findings shown in this paper clearly indicate that improved results are obtained by incorporating the equality constrains using a XY Hamiltonian versus using a penalization approach. This is likely due to the reduction in the search space introduced to the model by the former, as compared to the latter. While the incorporation of equality constrains are already successfully incorporated using this framework, it is still unknown how to incorporate alternative constraints, such as inequality constrains or logic relationships between decision variables that commonly appear in similar contexts. Particularly important are constraints that prevent a phenomenon known as "sensor clustering" in the structure, where optimal configurations tend to group the sensors in certain parts of the structure. From a practical point of view, this is not desired as it prevents practitioners from analyzing all locations in the structural system, and therefore suitable constraints should be incorporated to prevent this clustering phenomena. To tackle these challenges, further research into alternative mixer Hamiltonians should take place, as this option avoids the incorporation of additional penalization hyperparameters into the model.

Finally, while the use of the ADAM optimizer in this paper shows promising results, enhanced results could be obtained using optimization routines specifically designed for the QAOA algorithm, that possibly exploit the physical constraints imposed by the Quantum Adiabatic Algorithm. In this case, a constrained optimization forming a particular ordering in the sets $\{\beta_i\}_{i=1}^{P}$ and $\{\gamma_i\}_{i=1}^{P}$seems like an attractive area for future development in the area.

It is worth noting that the aforementioned research paths share a common attribute – they can be explored using a quantum simulator environment, without requiring the availability of a quantum computer. Therefore, they can be tackled in parallel to the development of quantum-capable hardware. As a final comment, we want to remark that the aforementioned challenges are independent to the fact that a quantum computer simulator was used to produce the results presented in this paper. Prior literature in the area commonly suggests that the performance of quantum algorithm will be improved once quantum-capable hardware becomes available at a large scale. We disagree with this notion. Improved results should not be expected by simply exchanging the quantum simulator environment for an execution process on quantum hardware. Indeed, a quantum computing simulation framework assumes an ideal, noise-free quantum computer and, therefore, measurement noise is not considered. This represents ideal execution conditions that are not likely to be achieved in the immediate future. As such, the challenges encountered in quantum computing algorithms nowadays should not be thought of as challenges produced by the lack of quantum hardware, but challenges related to the algorithms themselves. Consequently, we firmly believe that addressing these "algorithmic" challenges by early testing of quantum approaches in different disciplines, including Structural Health Monitoring, is just as critical as the development of large-scale, noise-free quantum hardware for the wide-spread adoption of quantum computing algorithms in practical engineering settings.

## Data availability

The code repository and associated data used to support the findings of this study have been deposited in a GitHub repository which can be accessed at [41].

## Conflicts of interest

The authors declare that there is no conflict of interest regarding the publication of this article.

## Funding statement

This research was conducted without specific external funding. Both authors were affiliated with the University of California, Los Angeles (UCLA) during the course of this work.

## Acknowledgements

A preprint has previously been published at [22].